\documentclass[12pt]{article}%
\usepackage{amsfonts}
\usepackage{amsmath,amssymb}
\usepackage{graphicx}
\usepackage{amsmath}
\usepackage{amssymb}%
\pdfoutput=1
\makeatletter

\@addtoreset{equation}{section}
\makeatother
\newcommand{\beq}{\begin{equation}}
\newcommand{\eeq}{\end{equation}}

\begin{document}

\baselineskip=18pt
\baselineskip 0.7cm

\begin{titlepage}
\setcounter{page}{0}
\renewcommand{\thefootnote}{\fnsymbol{footnote}}
\begin{center}
{\LARGE \bf
The Geometry of Supermanifolds and \\
New Supersymmetric Actions
\vskip .5cm
}
{\large
L. Castellani$^{~a,b,}$\footnote{leonardo.castellani@mfn.unipmn.it},
R. Catenacci$^{~a,c,}$\footnote{roberto.catenacci@mfn.unipmn.it},
and
P.A. Grassi$^{~a,b,}$\footnote{pietro.grassi@mfn.unipmn.it}\,.
\medskip
}
\vskip 0.5cm
{
\small\it
\centerline{$^{(a)}$ Dipartimento di Scienze e Innovazione Tecnologica, Universit\`a del Piemonte Orientale} }
\centerline{\it Viale T. Michel, 11, 15121 Alessandria, Italy}
\medskip
\centerline{$^{(b)}$ {\it
INFN, Sezione di Torino, via P. Giuria 1, 10125 Torino} }
\centerline{$^{(c)}$ {\it
Gruppo Nazionale di Fisica Matematica, INdAM, P.le Aldo Moro 5, 00185 Roma} }
\vskip  .3cm
\medskip
\end{center}
\smallskip
\centerline{{\bf Abstract}}
\medskip
\noindent
{This is the first of two papers in which we construct the Hodge dual for supermanifolds by means of the Grassmannian Fourier
transform of superforms. In this paper we introduce the fundamental concepts and a method for computing Hodge duals in simple cases. 
We refer to a subsequent publication \cite{lavoro3}
for a more general approach and the required mathematical details.}
In the case of supermanifolds it is known that superforms are not sufficient
to construct a consistent integration theory and that {\it integral forms}
are needed. They are distribution-like forms which can be integrated on supermanifolds
as a top form can be integrated on a conventional manifold.
In our construction of the Hodge dual of superforms they arise naturally.
The compatibility between Hodge duality and supersymmetry is exploited
and applied to several examples.
We define the irreducible representations of supersymmetry in terms
of integral and super forms in a new way which
can be easily generalised to several models in different dimensions.
The construction of supersymmetric
actions based on the Hodge duality is presented and new supersymmetric actions with higher derivative
terms are found. These terms are required by the invertibility of the Hodge operator. 

\vskip  .5cm
\noindent

{\today}
\end{titlepage}
\setcounter{page}{1}

\vfill
\eject
\tableofcontents

\vfill
\eject
\newpage\setcounter{footnote}{0} \newpage\setcounter{footnote}{0}

\section{Introduction}

In a series of previous papers \cite{Catenacci:2010cs,Catenacci,Castellani:2014goa} we
discussed several aspects of integral forms and their applications
\cite{Witten:2012bg,Berkovits:2004px}. Nonetheless, some of the issues are still only
partially understood and clarified, for example the generalization of the
usual Hodge dual was not clearly identified. Therefore we decided to use a
different point of view to study integral forms through the introduction
of an integral representation of integral forms. In this paper we face the
problem of constructing a generalization of the usual Hodge duality by means
of an integral representation of the Hodge operator. In this formalism the
integral forms naturally arise. The introduction of the Hodge operator is
relevant for constructing actions and for defining self-dual forms, and
reveals new features we study in the present paper and that will be
pursued in forthcoming publications.

The superspace techniques are well understood and used in quantum field theory
and string theory (see {\cite{Wess:1992cp,Gates:1983nr}}). They provide a very
powerful method to deal with supersymmetric multiplets and to write
supersymmetric quantities such as actions, currents, operators, vertex
operators, correlators and so on. This is based on the extension of the usual
space $\mathbb{R}^{n}$ obtained by adding to the bosonic coordinates $x^{i}$
some fermionic coordinates $\theta^{\alpha}$. One can take this construction
more seriously and extend the concept of superspace to a curved supermanifold
 which is locally homeomorphic to superspace.
Contextually, the many of the geometric structures which can be defined
for a conventional bosonic manifold can be rephrased in the new framework. For
example, the supermanifolds have a tangent bundle (generated by commuting and
anticommuting vector fields) and an exterior bundle. Therefore, one expects
that also the geometric theory of integration on manifolds could be exported
as it stands. Unfortunately, this is not so straightforward since top
\textit{superforms} do not exist. Before clarifying this point, we have to
declare what we mean by a superform. Even though there is no unanimous
agreement, we call \textit{superforms} the sections of the exterior
bundle constructed through generalized wedge products of the basic 1-forms
$dx^{i}$ and $d\theta^{\alpha}$ (that reduces to the ordinary wedge product
when only the basic 1-forms $dx^{i}$ are involved). The sets of fixed degree
superforms are modules over the ring of superfunctions $f(x,\theta)$. However,
while for the bosonic 1-forms $dx^{i}$ the usual rules are still valid, for
the fermionic 1-forms $d\theta^{\alpha}$ the graded Leibniz rule for $d$ (w.r.t. wedge product) has to be accompanied by the
anticommuting properties of fermionic variables, and this implies that a
fermionic 1-form commutes with itself and with all other forms. Thus, there
is no upper bound on the length of the usual exterior $d$-complex. To overcome
this problem, one needs to extend the concept of superforms including also
distributional-like forms, known as \textit{integral forms} \cite{Voronov2,integ}.
With a suitable extension of the $d$ differential they do form a complex with an upper bound,
and they can be used to define a meaningful geometric integration theory for
forms on supermanifolds. Clearly, this does not rely on any choice of
additional structure on the supermanifold (\textit{i.e.} complex structure,
Riemannian metric, connection, etc...) and it automatically gives a
diffeomorphism invariant theory of integration. This is important
for guaranteering parametrization-independence of the results, with the add-on
of the invariance under local supersymmetry as a part of the reparametrization  invariance
of the entire supermanifold. The details of this construction are contained in
several papers \cite{Catenacci:2010cs,Castellani:2014goa}  and we will give in the following only a
short review of the most important points.

In a supermanifold $\mathcal{M}^{(n|m)}$ with $n$ bosonic dimensions and $m$
fermionic dimensions, there is a Poincar\'{e} type duality between forms of
the differential complexes. In that respect, we have to use the complete set
of forms comprehending both superforms and integral forms. It can be shown
that (when finitely generated) there is a match between the dimensions of the
modules of forms involved in this duality. Then, as in the conventional
framework, we are motivated to establish a map between them, conventionally
denoted as \textit{Hodge duality}. In order to be a proper generalization of
the usual Hodge dual, this map has to be involutive, which implies its
invertibility (as discussed in the forthcoming section, the lack of
invertibility for a generic linear map leads to problems). We first show
that the conventional Hodge duality for a bosonic manifold can be constructed
using a ``partial" Fourier transform of differential forms (for a ``complete'' Fourier transform
see also \cite{Berkovits:2009gi,Kalkman:1993zp}). Then we extend it to superforms.
By ``partial" we mean a Fourier transformation only of the
differentials $dx$ and $d\theta$, leaving untouched the coordinates $x$ and
$\theta$ and hence the components of the superform. To compute the general
form of the Hodge duality we start with the case of a standard constant
diagonal metric. For a slightly more general metric, we consider a transformation of the
basic 1-forms that diagonalizes it and afterwards rewrite the standard
Hodge dual in terms of the original differentials. This is equivalent to
passing from the holonomic to the anholonomic basis with a Cartan super frame
(supervielbein). 
Finally, we show that the compatibility with supersymmetry
constrains the form of the supervielbein and that the supersymmetric-invariant
variables are indeed those for which the Hodge operator is diagonal.
As an example, we work out completely a very simple one dimensional model.

{The definition of the super Hodge dual can be extended to the general metrics needed in physical applications.
We refer to the paper \cite{lavoro3} for the generalization and more mathematical details.}

With the definition of the Hodge operator we have a new way to build new
Lagrangians and the corresponding actions in terms of superforms and their
differentials. For that purpose, we first give some examples in the case of a
three dimensional bosonic manifold with two additional fermionic coordinates.
This is one of the simplest supermanifolds, but displays several features of
higher dimensional models. In particular, there are different types of
supermultiplets such as the scalar superfield, the vector superfield and
current superfield. They can be formulated in the present new geometrical
framework and their corresponding actions can be built. The interesting result
is that the action only partially coincides with the conventional result, since
there are additional higher derivative terms required by the invertibility of the
Hodge dual operation. Moving from three to
four dimensions, we find new examples of multiplets and for them
we give a geometrical definition. We construct the actions as integrals on the
corresponding supermanifold.

\subsection{Motivations and some old results}

In this section we briefly outline the motivations of our study describing some old results and observations regarding the
problems encountered in building Lagrangians and actions on supermanifolds. We anticipate some notations and concepts
that will be described and explained in the forthcoming sections.

In previous works (see for example \cite{Catenacci:2010cs})
we have seen that there is a Poincar\'e duality among 
forms $\Omega^{(p|q)}({\cal M}^{(n|m)})$  on the supermanifold ${\cal M}^{(n|m)}$
expressed by the relation
$$
\Omega^{(p|0)} \longleftrightarrow  \Omega^{(n-p|m)}\,.
$$
Here the numbers $p$ and $q$ respectively denote the {\it form degree}
(the usual form degree, which in the case of integral forms could also be negative)
and the {\it picture number} (taking into account the number of Dirac delta
forms of type $\delta(d\theta^\alpha)$ where $d\theta^\alpha$ is the fundamental 1-form associated to
the coordinates $\theta^\alpha$ of the supermanifold ${\cal M}^{(n|m)}$ with $\alpha=1, \dots, m$).

Let us set the stage by
considering the N=1 Wess-Zumino model in three dimensions. The
${\cal M}^{(3|2)}$ supermanifold is locally homeomorphic to $\mathbb{R}^{(3|2)}$
parametrised by 3 bosonic coordinates $x^m$ and 2 fermionic coordinates $\theta^\alpha$.
A top form $\Omega_{top}$ is an integral form belonging to $\Omega^{(3|2)}$ (which is one dimensional)
\begin{equation}\label{topform}
J_{top} = h(x,\theta) d^3x \delta^2(d\theta)\,,
\end{equation}
where $h(x,\theta)$ is a superfield and $\delta^2(d\theta) =  \delta(d\theta^\alpha) \epsilon^{\alpha\beta} \delta(d\theta^\beta)$.
Such a form can be integrated on the supermanifold as discussed in \cite{Castellani:2014goa}.
If $h(x, \theta) = h_0(x) + h_\alpha(x) \theta^\alpha + h_2(x) \theta^2/2$ (where $\theta^2 = \theta^\alpha \epsilon_{\alpha\beta} \theta^\beta$
),
the integral of $J_{top}$ on the supermanifold ${\cal M}$ is given by
\begin{equation}\label{forA}
\int_{\cal M} J_{top} = \int_{M} \epsilon^{\alpha\beta} D_\alpha D_\beta
\left. h(x, \theta)\right|_{\theta=0} d^3x = \int_M h_2(x) d^3x
\end{equation}
where $M$ is the bosonic submanifold of ${\cal M}$
and $D_\alpha = \frac{\partial}{\partial \theta^\alpha}$.
There are three ways to build an action using the forms $\Omega^{(p|q)}$.

The first one is by considering a Lagrangian ${\cal L}(x,\theta)$ belonging to $\Omega^{(0|0)}$
(a function on the supermanifold) and then map it to an integral form of the type
$\Omega^{(3|2)}$ by introducing a linear application (which we ``improperly" call {\it Hodge operator})
\begin{equation}
{\cal L} \in \Omega^{(0|0)} \rightarrow \star {\cal L} \in \Omega^{(3|2)}\,.
\end{equation}
For that we need to establish what is the Hodge dual of the generator of $\Omega^{(0|0)}$, namely
we need to know what is $\star 1$. We assume that
\begin{equation}
\star 1 = h(x,\theta) d^3x \delta^2(d\theta)
\end{equation}
 so that $\int_{\cal M} \star 1 = \int_M h_2(x) d^3x $.
 Then, we find
  \begin{equation}
S = \int_{\cal M} \star {\cal L} = \int_{\cal M} {\cal L}(x,\theta) \,  h(x, \theta)  d^3x \delta^2(d\theta)=
\end{equation}
$$
=\int_{M} \Big(
h_0 \left. D^2 {\cal L}(x, \theta)\right|_{\theta=0} +
2 h_\alpha(x) \left. D^\alpha {\cal L}(x, \theta)\right|_{\theta=0} +
h_2(x) \left. {\cal L}(x, \theta)\right|_{\theta=0} \Big)
d^3x\,.
$$
 We immediately notice that the $\star$ operation is singular if $h_0(x)$ and $h_\alpha(x)$ vanish,
 since the relevant part of action is only that for $\theta=0$ and we can shift it by any
 $\theta$-dependent term without modifying the action.
 This means that the equations of motion derived in this case are the $\theta=0$ projected equations.

For the second way, we start from a superform  ${\cal L} \in \Omega^{(3|0)}$, and then map it to
the space $\Omega^{(3|2)}$, by means of the Picture Changing Operator
$Y^2 = \theta^2 \delta^2({d\theta})$. This operator has been discussed in \cite{Catenacci:2010cs}
where it is shown that it corresponds to a generator of a non-trivial cohomology class and it
can be used to relate differential forms of the type $\Omega^{(p|0)}$ to
differential forms of the type $\Omega^{(p|2)}$ with maximum number of Dirac delta's. It is
also shown that $Y^2$ maps the cohomology class $H^{(p|0)}_d$ onto $H^{(p|2)}_d$.
So, given ${\cal L}$, we can define an integral form of the type (\ref{topform})
as follows
\begin{equation}
{\cal L} \in \Omega^{(3|0)} \longrightarrow   Y^2  {\cal L} \in \Omega^{(3|2)}\,.
\end{equation}
A $3$-superform can be decomposed into pieces
\begin{equation}
{\cal L} = {\cal L}_{[mnp]} dx^m dx^n dx^p  + {\cal L}_{\alpha[mn]} d\theta^\alpha dx^m dx^n + \dots +  {\cal L}_{(\alpha \beta \gamma)} d\theta^\alpha
d\theta^\beta d\theta^\gamma\,,
\end{equation}
where the coefficients ${\cal L}_{[mnp]} = \epsilon_{mnp} {\cal L}_0, {\cal L}_{\alpha[mn]}, {\cal L}_{(\alpha \beta) m}, {\cal L}_{(\alpha \beta \gamma)}$
are superfields.
Thus, the action now reads
\begin{equation}
S = \int_{\cal M} Y^2 {\cal L} = \int_{\cal M}  \theta^2 \delta^2({d\theta}) {\cal L}  = \int_M  {\cal L}_0 (x, 0) d^3x\,,
\end{equation}
where  only the first coefficient of the superform survives and it is computed at $\theta =0$. In the present
computation the arbitrariness is even greater than before, ${\cal L}$ is defined up to any
superform which is proportional to $\theta$ or to a power of $d\theta$.

A third way is to construct the action by writing an integral form of the type (\ref{topform})  in terms of other forms.
Given a supefield $\Phi \in \Omega^{(0|0)}$, its (super)differential $d\Phi \in \Omega^{(1|0)}$ and
using the linear map as above we find $\star d\Phi \in \Omega^{(2|2)}$; then
we can define the Lagrangian as follows
\begin{equation}
{\cal L} = d\Phi \wedge \star d \Phi \in \Omega^{(3|2)}\,.
\end{equation}
Then, the action is an integral form
and it can be integrated on the supermanifold.
To compute the action, we must decompose the superfield $\Phi$
\begin{equation}\label{WZA}
\Phi = A + \psi^\alpha \theta_\alpha + F \theta^2/2 \,,
\end{equation}
where $A, \psi^\alpha, F$ are the component fields. Let us take the differential of $\Phi$
\begin{equation}\label{WZB}
d \Phi = \partial_m \Phi dx^m + \partial_\alpha \Phi d\theta^\alpha\,,
\end{equation}
Now, we write the linear map $d\Phi \longrightarrow \star d \Phi$ as follows
\begin{eqnarray}\label{HODa}
\star dx^m &=& G^{m n}(x,\theta) \epsilon^{n p q} dx^p  dx^q  \delta^2(d\theta) +  G^{m \alpha}(x,\theta) d^3x \iota_\alpha \delta^2(d\theta) \,, \\ \nonumber
\star \, d\theta^\alpha &=& G^{\alpha n}(x,\theta) \epsilon^{n p q} dx^p  dx^q  \delta^2(d\theta) +  G^{\alpha \beta}(x,\theta) d^3x  \iota_\beta \delta^2(d\theta) \,,
\end{eqnarray}
where $\iota_\alpha \delta^2(d\theta)$ is the derivative of the Dirac delta forms with respect to the argument
$d\theta^\alpha$ and it satisfies $d\theta^\alpha \iota_\beta \delta^2(d\theta) = - \delta^\alpha_\beta \delta^2(d\theta)$.
Notice that the 1-forms $dx^m, d\theta^\alpha$ belong to $\Omega^{(1|0)}$ and therefore the
"Hodge dual" should belong to $\Omega^{(2|2)}$ and it is easy to check that this space is generated by two elements.
Therefore, it is natural that the Hodge dual of $d\Phi$ is a combination of the two elements.
The entries of the supermatrix
\begin{equation}\label{HODb}
\mathbb{G} = \left(
\begin{array}{cc}
G^{m n}(x,\theta)  &   G^{m \beta}(x,\theta) \\
 G^{\alpha n}(x,\theta)  & G^{\alpha \beta}(x,\theta)
\end{array}
\right)
\end{equation}
are superfields.
Then, we have
\begin{eqnarray}\label{WZC}
\star d \Phi &=&  \partial_m \Phi \left(
G^{m n}\epsilon_{n p q} dx^p  dx^q  \delta^2(d\theta) +  G^{m \beta} d^3x  \iota_\beta \delta^2(d\theta)\right) \nonumber
\\
&+&
\partial_\alpha \Phi \left(  G^{\alpha n} \epsilon_{n p q} dx^p  dx^q  \delta^2(d\theta) +  G^{\alpha \beta}d^3x  \iota_\beta \delta^2(d\theta) \right)\,.
 \end{eqnarray}
Finally, we can compute
$$
d\Phi \wedge \star\, d\Phi =
$$
\begin{equation}\label{WZD}
= \Big( \partial_m \Phi G^{mn} \partial_n \Phi +  \partial_m \Phi G^{m\beta} \partial_\beta \Phi +
\partial_\alpha \Phi G^{\alpha m} \partial_m \Phi + \partial_\alpha \Phi G^{\alpha\beta} \partial_\beta \Phi \Big) \, d^3x  \delta^2(d\theta)
\end{equation}
and, hence, by integrating over $d\theta$ and over $\theta$ (by Berezin integral) we obtain
\begin{equation}\label{WZE}
\int_{\cal M} d\Phi \wedge \star d\Phi  =
\int_{ M}  d^3x  (\partial_m A \partial^m A + \psi^\alpha \gamma^m_{\alpha\beta} \partial_m \psi^\beta + F^2) \,
\end{equation}
by choosing
\begin{equation}\label{WZf}
\mathbb{G} = \left(
\begin{array}{cc}
G^{m n}(x,\theta)  &   G^{m \alpha}(x,\theta) \\
 G^{\beta n}(x,\theta)  & G^{\alpha \beta}(x,\theta)
\end{array}
\right) =
\left(\begin{array}{cc}
\eta^{m n} \theta^2  &   \gamma^{m \alpha \beta}\theta_\beta \\
\gamma^{n \alpha \beta}\theta_\alpha   & \epsilon^{\alpha \beta}
\end{array}
\right)
\end{equation}
where $\gamma^{m\alpha\beta}$ are the Dirac matrices in 3d.

Notice that the matrix $\mathbb{G}$ has non-vanishing superdeterminant (by suitable choice of the numerical factors), however it is proportional
to $\theta^2$ and therefore it cannot be inverted. So, in this way we have constructed an action principle which leads to
the correct equations of motion, but at the price of a non-invertible Hodge operator.

\section{Super Fourier Transforms}

In this section we present the theory of Fourier transforms in Grassmann
algebras and its generalizations to differential forms, super forms and
integral forms. This formalism will be used to define an invertible Hodge dual on supermanifolds.

The case of the Fourier transform of usual differential forms on
differentiable manifolds was described for example in \cite{Kalkman:1993zp}.
We will rephrase the formalism in such a way that it will
allow us to extend the Fourier transform to super and integral forms on supermanifolds.

These generalizations are then applied to define a Hodge dual for super and integral forms.

Appendices A and B contain some preliminary observations about the
use of Fourier transforms in the cohomology of superforms. This matter will be
expanded in a forthcoming publication.

\subsection{Fourier transform in Grassmann algebras}

We start, as usual, from the case of the real superspace $\mathbb{R}^{n|m}$
with $n$ bosonic ($x^{i},i=1,\dots,n$) and $m$ fermionic $(\theta^{\alpha
},\alpha=1,\dots,m)$ coordinates. We take a function $f(x,\theta)$ in
$\mathbb{R}^{n|m}$ with values in the real algebra generated by $1$ and by the
anticommuting variables, and we expand $f$ as a polynomial in the variables
$\theta:$%
\begin{equation}
f(x,\theta)=f_{0}(x)+...+f_{m}(x)\theta^{1}...\theta^{m}\,.%
\end{equation}
Recall that if the real function $f_{m}(x)$ is integrable in some sense in
$\mathbb{R}^{n},$ the Berezin integral of $f(x,\theta)$ is defined as:%
\begin{equation}
\int_{\mathbb{R}^{n|m}}f(x,\theta)[d^{n}xd^{m}\theta]=\int_{\mathbb{R}^{n}\,.%
}f_{m}(x)d^{n}x
\end{equation}
Here and in the following we use the notations of \cite{Castellani:2014goa}.

To define super Fourier transforms we start from the complex vector space $V$
spanned by the $\theta^{\alpha}$:%

\[
V=Span_{\mathbb{C}}\{\theta^{\alpha},\alpha=1,\dots,m\}
\]
and we denote as usual by%
\[
\bigwedge(V)=\sum_{p=0}^{m}\bigwedge^{p}(V)
\]
the corresponding {\it complex} Grassman algebra of dimension $2^{m}$.

If $F(\mathbb{R}^{n})$ is some suitable functional space of real or complex valued
functions in $\mathbb{R}^{n},$ the functions $f(x,\theta)$ in $\mathbb{R}%
^{n|m}$ are elements of $F(\mathbb{R}^{n})\otimes\bigwedge(V).$

Berezin integration restricted to $\bigwedge(V)$ is simply a linear map
$\int(\cdot)[d^{m}\theta]$ from $\bigwedge(V)$ to $\mathbb{C}$ that is zero on
all elements other than the product $\theta^{1}...\theta^{m}\in\bigwedge
^{m}(V)$ 
\begin{equation}
\int\theta^{1}...\theta^{m}[d^{m}\theta]=1\,.
\end{equation}
This can be extended to a linear map $\int(\cdot)[d^{m}\theta]$ from
$\bigwedge(V^{\ast})\otimes\bigwedge(V)$ to $\bigwedge(V^{\ast})$ where
$V^{\ast}$ is the dual space of $V$. If $\psi\in\bigwedge(V^{\ast})$ we simply
define:%
\begin{equation}
\int\psi\otimes\theta^{1}...\theta^{m}[d^{m}\theta]=\psi\,.
\end{equation}

Denoting with $\{\psi_{\alpha},\alpha=1,\dots,m\}$ the dual basis of the basis
$\{\theta^{\alpha},\alpha=1,\dots,m\},$ for every $\omega\in$ $\bigwedge(V)$
the Fourier transform $\mathcal{F}$ is defined by:%
\begin{equation}
\mathcal{F}(\omega)(\psi)=\int\omega(\theta)e^{i\psi_{\alpha}\otimes
\theta^{\alpha}}[d^{m}\theta]\in\bigwedge(V^{\ast}) \label{fourier1}\,.%
\end{equation}
We will denote also by $\mathcal{F}$ the (anti)transform of $\eta\in$
$\bigwedge(V^{\ast})$:
\begin{equation}
\mathcal{F}(\eta)(\theta)=\int\eta(\psi)e^{i\theta^{\alpha}\otimes\psi
_{\alpha}}[d^{m}\psi] \label{antifourier}\,.%
\end{equation}
Recall now that for
$\mathbb{Z}_{2}-$ graded algebras $A$ and $B,$ the tensor product must be
defined in such a way that the natural isomorphism $A\otimes B\simeq B\otimes
A$ holds with a sign: for $a\in A$ and $b\in B$ we have:
\begin{equation}
a\otimes b\longrightarrow(-1)^{p(a)p(b)}b\otimes a
\end{equation}
(where $p(a)$ and $p(b)$ denote the $\mathbb{Z}_{2}$-parity of the elements
$a$ and $b).$
The exponential series is defined recalling also that
if $\mathcal{A}$ and $\mathcal{B}$ are two $\mathbb{Z}_{2}$-graded algebras
with products $\cdot_{\mathcal{A}}$and $\cdot_{\mathcal{B}}$, the
$\mathbb{Z}_{2}$-graded tensor product $\mathcal{A}\otimes\mathcal{B}$ is a
$\mathbb{Z}_{2}$-graded algebra with the product given by (for homogeneous
elements);%
\[
(a\otimes b)\cdot_{\mathcal{A}\otimes\mathcal{B}}(a^{\prime}\otimes b^{\prime
})=(-1)^{\left\vert a^{\prime}\right\vert \left\vert b\right\vert }%
a\cdot_{\mathcal{A}}a^{\prime}\otimes b\cdot_{\mathcal{B}}b^{\prime}%
\]
In the following the tensor product symbol will be omitted.

Note that the exponential series stops at the $m^{th}$ power and that the
factor $i$ in the exponential is here only for ``aesthetic reasons" and it is
of no importance for the existence of the fermionic integral.

As a simple example let us consider a two dimensional $V$ generated over
$\mathbb{C}$ by $\left\{  \theta^{1},\theta^{2}\right\}  .$ We take
$\omega=a+b\theta^{1}+c\theta^{2}+d\theta^{1}\theta^{2}\in$ $\bigwedge(V)$ and
compute%
\[
e^{i\left(  \psi_{1}\theta^{1}+\psi_{2}\theta^{2}\right)  }=1+i\psi_{1}%
\theta^{1}+i\psi_{2}\theta^{2}+\psi_{1}\psi_{2}\theta^{1}\theta^{2}\,.%
\]
We find:%
\[
\mathcal{F}(\omega)=\int\left(  a+b\theta^{1}+c\theta^{2}+d\theta^{1}%
\theta^{2}\right)  \left(  1+i\psi_{1}\theta^{1}+i\psi_{2}\theta^{2}+\psi
_{1}\psi_{2}\theta^{1}\theta^{2}\right)  [d^{2}\theta]=d+ic\psi_{1}-ib\psi
_{2}+a\psi_{1}\psi_{2}\,.%
\]
Note that $\mathcal{F}$ maps $\bigwedge^{p}(V)$ in $\bigwedge^{m-p}(V^{\ast
}).$

This definition shares many important properties with the usual case, for
example one has (this will be proved in the following, see the formula
(\ref{quadratodifourier})){: }%
\begin{equation}
\mathcal{F}^{2}=(i)^{m^{2}}1_{\bigwedge(V)}%
\end{equation}
Hence, if $m$ is even, as is usual in many physical applications:
\begin{equation}
\mathcal{F}^{2}=1_{\bigwedge(V)}%
\end{equation}
In $\bigwedge(V)$ there is a convolution product. For $\omega$ and $\eta\in$
$\bigwedge(V)$ one defines:
\begin{equation}
\left(  \omega\ast\eta\right)  \left(  \theta\right)  =\int\omega
(\theta^{\prime})\eta(\theta-\theta^{\prime})[d^{m}\theta^{\prime}]
\label{convo1}%
\end{equation}
This convolution in $\bigwedge(V)$ obeys the usual rules:%
\begin{subequations}
\begin{align}
\mathcal{F}(\omega\ast\eta)  &  =\mathcal{F}(\omega)\mathcal{F}(\eta
)\label{convo2}\\
\mathcal{F}(\omega\eta)  &  =\mathcal{F}(\omega)\ast\mathcal{F}(\eta)
\label{convo3}%
\end{align}
Taking for example, $\omega=1+\theta^{1}$ and $\eta=1+\theta^{2}$, we
have
\begin{equation}
\omega\ast\eta(\theta)=\int\left(  1+\theta^{\prime1}\right)  \left(
1+\theta^{2}-\theta^{\prime2}\right)  [d^{2}\theta^{\prime}]=-1
\end{equation}
and the (\ref{convo2}) and (\ref{convo3}) are immediately verified.

One can now combine the definition (\ref{fourier1}) with the usual Fourier
transform in order to obtain the Fourier transform of the functions
$f(x,\theta)$ in $\mathbb{R}^{n|m}$. We are not interested here in analytic
subtelties and we limit ourselves to some "suitable" functional space (for
example the space of fast decreasing functions) for the "component functions"
of $f(x,\theta)=f_{0}(x)+...+f_{1\dots m}(x)\theta^{1}...\theta^{m}.$ In the
following we will also consider its dual space of tempered distributions.

If the $y_{i}$ are variables dual to the $x^{i}$ one can define:%
\end{subequations}
\begin{equation}
\mathcal{F}(f)=\int_{\mathbb{R}^{n|m}}f(x,\theta)e^{i(y_{i}x^{i}+\psi_{\alpha
}\theta^{\alpha})}[d^{n}xd^{m}\theta] \label{fourier2}%
\end{equation}
As a simple example let us consider again $\mathbb{R}^{1|2}.$ We have
$f(x,\theta)=f_{0}(x)+f_{1}(x)\theta^{1}+f_{2}(x)\theta^{2}+f_{12}%
(x)\theta^{1}\theta^{2}$ and hence:%
\[
\mathcal{F}(f)\left(  y,\psi\right)  =\widehat{f_{12}}(y)+i\widehat{f_{2}%
}(y)\psi_{1}-i\widehat{f_{1}}(y)\psi_{2}+\widehat{f_{0}}(y)\psi_{1}\psi_{2}%
\]
Where $\widehat{f}(y)$ denotes the usual Fourier transform of the function
$f(x)$. In the following we will denote $\widetilde{g}(x)$ the usual
antitransform of the function $g(y)$.

Note that we can extend the definition (\ref{fourier2}) to more general
$f(x,\theta)$ (with component functions not rapidly decreasing). For example:%
\begin{equation}
\int_{\mathbb{R}^{1|1}}e^{i(yx+\psi\theta)}[dxd\theta]=i\delta(y)\psi
\end{equation}
Similar expressions hold in higher dimensions.

The convolution in $\bigwedge(V)$ described above can be extended to produce a
convolution in $\mathbb{R}^{n|m}:$%
\begin{equation}
\left(  f\ast g\right)  (x,\theta)=\int_{\mathbb{R}^{n|m}}f(x^{\prime}%
,\theta^{\prime})g(x-x^{\prime},\theta-\theta^{\prime})[d^{n}xd^{m}\theta]
\label{convo4}%
\end{equation}

\subsection{Fourier transform of differential forms}

The formalism described above can be used to define the Fourier transform of a
differential form. For this we exploit the similarity between the Berezin
integral and the usual integral of a differential form, that we now briefly recall.

Denoting by $M$ a differentiable manifold with dimension $n$, we define the
exterior bundle $\Omega^{\bullet}(M)=\sum_{p=0}^{n}\bigwedge^{p}(M)$ as the
direct sum of $\bigwedge^{p}(M)$ (sometimes denoted also by $\Omega^{p}(M)$).
A section $\omega$ of $\Omega^{\bullet}(M)$ can be written locally as
\begin{equation}
\omega=\sum_{p=0}^{n}\omega_{i_{1}\dots i_{p}}(x)dx^{i_{1}}\wedge\dots\wedge
dx^{i_{p}} \label{inA}%
\end{equation}
where the coefficients $\omega_{i_{1}\dots i_{p}}(x)$ $(i_{1}<...<i_{p})$ are
functions on $M$ and repeated indices are summed. The integral of $\omega$ is
defined as:
\begin{equation}
I[\omega]=\int_{M}\omega=\int_{M}\omega_{1...n}(x)\,d^{n}x\,, \label{inB}%
\end{equation}
suggesting a relation between the integration theory of forms and the Berezin
integral, that can be exploited by considering every $1$-form $dx^{i}$ as an
abstract Grassmann variable. A section \ $\omega$ of ${\Omega^{\bullet}(}M{)}$
is viewed locally as a function on a supermanifold $\mathcal{M}$ of dimension
$n|n$ with local coordinates $(x^{i},dx^{i}):$
\begin{equation}
\omega(x,dx)=\sum_{p=0}^{n}\omega_{i_{1}\dots i_{p}}(x)dx^{i_{1}}\dots
dx^{i_{p}}\,; \label{inAA}%
\end{equation}
such functions are polynomials in $dx^{i}$. Supposing now that the form
$\omega$ is integrable we see that the Berezin integral \textquotedblleft
selects" the top degree component of the form:%
\begin{equation}
\int_{\mathcal{M}}\omega(x,dx)[d^{n}xd^{n}\left(  dx\right)  ]=\int_{M}%
\omega\label{bereform}%
\end{equation}

\bigskip With this interpretation (and denoting $y$ and $dy$ the dual
variables) we can directly apply to (\ref{bereform}) to define the Fourier
transform of a differential form in $\mathbb{R}^{n}:$
\begin{equation}
\mathcal{F}(\omega)\left(  y,dy\right)  =\int_{\mathbb{R}^{n|n}}%
\omega(x,dx)e^{i(y_{i}x^{i}+dy_{i}dx^{i})}[d^{n}xd^{n}\left(  dx\right)  ]
\label{fourier4}%
\end{equation}
As an example consider a two-form $\omega$ in $\mathbb{R}^3$, that is $\omega = \sum_{i_1 i_2}
\omega_{i_1 i_2} (x) dx^{i_1}\wedge dx^{i_2}$.
Its Fourier transform is given by
\begin{equation}\label{FF}
{\cal F}(\omega) =
\int_{\mathbb{R}^{3|3}}%
\omega_{i_1 i_2} (x) dx^{i_1}\wedge dx^{i_2} e^{i(y_{i}x^{i}+dy_{i}dx^{i})}[d^{n}xd^{n}\left(  dx\right)  ]
\end{equation}
$$
= i \left( \widehat\omega_{12} dy_3 - \widehat\omega_{13} dy_2 + \widehat\omega_{23} dy_1\right)
$$
where $\widehat\omega_{i_1 i_2}$ is the usual Fourier transform of the functions
$\omega_{i_1 i_2}(x)$.

\subsection{Fourier transform of super and integral forms}

We denote now by $\mathcal{M}$ a supermanifold of dimension $n|m$ with
coordinates $(x^{i},\theta^{\alpha})$ (with $i=1,\dots,n$ and $\alpha
=1,\dots,m$) and we consider the \textquotedblleft exterior" bundle ${\Omega
}^{\bullet}(\mathcal{M})$ as the formal direct sum of bundles of fixed degree
forms. The local coordinates in the total space of this bundle are
$(x^{i},d\theta^{\alpha},dx^{j},\theta^{\beta}),$ where $\left(  x^{i}%
,d\theta^{\alpha}\right)  $ are bosonic and $\left(  dx^{j},\theta^{\beta
}\right)  $ fermionic. In contrast to the pure bosonic case, a top form does
not exist because the $1-$ forms of the type $d\theta^{\alpha}$ commute among
themselves $d\theta^{\alpha}\wedge d\theta^{\beta}=d\theta^{\beta}\wedge
d\theta^{\alpha}$. Then we can consider {\it superforms} of any degree (the
formal infinite sum is written here just to remind that we can have
homogeneous superforms of any fixed degree):
\begin{equation}
\omega(x,\theta,dx,d\theta)=\sum_{p=0}^{n}\sum_{l=0}^{\infty}\omega_{\lbrack
i_{1}\dots i_{p}](\alpha_{1}\dots\alpha_{l})}(x,\theta)dx^{i_{1}}...dx^{i_{p}%
}d\theta^{\alpha_{1}}\dots d\theta^{\alpha_{l}} \label{inG}%
\end{equation}

where the coefficients $\omega_{\lbrack i_{1}\dots i_{p}](\alpha_{1}%
\dots\alpha_{l})}(x,\theta)$ are functions on the supermanifold $\mathcal{M}$
with the first $1\dots p$ indices antisymmetrized and the last $1\dots l$
symmetrized. The component functions $\omega_{\lbrack i_{1}\dots i_{p}%
](\alpha_{1}\dots\alpha_{l})}(x,\theta)$ are polynomial expressions in the
$\theta^{\alpha}$ and their coefficients are functions of $x^{i}$ only.

It is clear now that {\it we cannot integrate a generic }$\omega\left(
x,\theta,dx,d\theta\right)  $ mainly because we do not have yet a general
definition of integration with respect to the $d\theta$ variables (we shall
return to this crucial point at the end of this paragraph). Moreover, suppose
that some integrability conditions are satisfied with respect to the $x$
variables; the integrals over $dx$ and $\theta$ (being Berezin integrals) pose
no further problem but, if $\omega\left(  x,\theta,dx,d\theta\right)  $ has a
polynomial dependence in the (bosonic) variables $d\theta,$ the integral,
however defined, "diverges". We need a sort of formal algebraic integration
also for the $d\theta$ variables.

In order to do so one introduces the Dirac's ``distributions" $\delta\left(
d\theta^{\alpha}\right)  $. The distributions $\delta(d\theta^{\alpha})$ have
most of the usual properties of the Dirac delta function $\delta(x)$, but, as
described at the end of this paragraph, one must impose:
\begin{equation}
\delta(d\theta^{\alpha})\delta(d\theta^{\beta})=-\delta(d\theta^{\beta}%
)\delta(d\theta^{\alpha}) \label{inH}%
\end{equation}
Therefore, the product $\delta^{m}(d\theta)\equiv\prod_{\alpha=1}^{m}%
\delta(d\theta^{\alpha})$ of all Dirac's delta functions (that we will call
also delta forms) serves as a \textquotedblleft top form".

One can then integrate the objects $\omega\left(  x,\theta,dx,d\theta\right)
$ provided that they depend on the $d\theta$ only through the product of {\it all }the
distributions $\delta\left(  d\theta^{\alpha}\right)  $. This solves the
problem of the divergences in the $d\theta^{\alpha}$ variables because
$\int\delta\left(  d\theta^{\alpha}\right)  [d\left(  d\theta^{\alpha}\right)
]=1.$

A {\it pseudoform} $\omega^{(p|q)}$ belonging to $\Omega^{(p|q)}%
(\mathcal{M})$ is characterised by two indices $(p|q)$: the first index is the
usual form degree and the second one is the {\it picture} number which
counts the number of delta forms (and derivatives of delta forms, see below).

A pseudoform reads:
\begin{equation}
{\omega^{(p|q)}=\sum_{r=0}^{p}\omega_{\lbrack i_{1}\dots i_{r}](\alpha
_{r+1}\dots\alpha_{p})[\beta_{1}\dots\beta_{q}]}(x,\theta)dx^{i_{1}}\dots
{}dx^{i_{r}}d\theta^{\alpha_{r+1}}\dots{}d\theta^{\alpha_{p}}\delta
(d\theta^{\beta_{1}})\dots\delta(d\theta^{\beta_{q}})} \label{integralform}%
\end{equation}
with $\omega_{\lbrack i_{1}\dots i_{r}](\alpha_{r+1}\dots\alpha_{p})[\beta
_{1}\dots\beta_{q}]}(x,\theta)$ superfields.

An {\it integral form} is a pseudoform without $d\theta$ components . Note
however that in the literature there is no complete agreement on these definitions.

The $d\theta^{\alpha}$ appearing in the product and those appearing in the
delta functions are reorganised respecting the rule $d\theta^{\alpha}%
\delta(d\theta^{\beta})=0$ if $\alpha=\beta$. We see that if the number of
delta's is equal to the fermionic dimension of the space no $d\theta$ can
appear; if moreover the number of the $dx$ is equal to the bosonic dimension
the form $($of type ${\omega^{(n|m)})}$ is an {\it integral top form, }the
only objects we can integrate on $\mathcal{M}$. It would seem that integrals
on supermanifolds the $d\theta$-components of the integrands are ruled out. However,
${\omega^{(p|q)}}$ as written above is not yet the most generic
pseudoform, since we could have added the derivatives of delta forms (and they
indeed turn out to be unavoidable and play an important role). They act by
reducing the form degree (so we can have negative degree pseudoforms)
according to the rule $d\theta^{\alpha}\delta^{\prime}(d\theta^{\alpha
})=-\delta(d\theta^{\alpha})$, where $\delta^{\prime}(x)$ is the first
derivative of the delta function with respect to its variable. (We denote also
by $\delta^{(p)}(x)$ the p-derivative). This observation is fundamental to establish
the isomorphism between the space of superforms (at a given form degree)
and the space of integral forms, namely $\Omega^{(p|0)}(\mathcal{M})$ and $\Omega^{(n-p|m)}(\mathcal {M})$.

In general, if $\omega$ is an integral form in $\Omega^{\bullet}(\mathcal{M}%
)$, its integral on the supermanifold is defined (in analogy with the Berezin
integral for bosonic forms) as follows:
\begin{equation}
\int_{\mathcal{M}}\omega\equiv\int_{\mathcal{M}}\omega_{\lbrack1...n][1...m]}%
(x,\theta)[d^{n}x\,d^{m}\theta] \label{inLAA}%
\end{equation}
where the last integral over $\mathcal{M}$ is the usual Riemann-Lebesgue
integral over the coordinates $x^{i}$ (if it exists) and the Berezin integral
over the coordinates $\theta^{\alpha}$. The expressions $\omega_{\lbrack
i_{1}\dots i_{n}][\beta_{1}\dots\beta_{m}]}(x,\theta)$ denote those components
of the pseudoform (\ref{integralform}) with no symmetric indices.

For the Fourier transforms we introduce dual variables as follows:%
\begin{align*}
y  &  \longleftrightarrow x\text{ (bosonic)}\\
\psi &  \longleftrightarrow\theta\text{ (fermionic)}\\
b  &  \longleftrightarrow d\theta\text{ (bosonic)} \\
\eta &  \longleftrightarrow dx\text{ (fermionic)}
\end{align*}

We define the Fourier transform of a superform $\omega$ in $\mathbb{R}^{n|m}$
as:%
\begin{equation}
\mathcal{F}(\omega)=\int_{\mathbb{R}^{n+m|n+m}}\omega(x,\theta,dx,d\theta
)e^{i(yx+\psi\theta+\eta dx+bd\theta)}[d^{n}xd^{m}\theta d^{n}\left(
dx\right)  d^{m}\left(  d\theta\right)  ] \label{fourier5}%
\end{equation}
where the functional dependence for the $\omega(x,\theta,dx,d\theta)$ that we
will consider is, for example, rapidly decreasing in the $x$ variables or,
more generally, tempered distributions in $x;$ polynomial in $\theta$ and
$dx,$ and depending on the $d\theta$ variables only through a product of
Dirac's delta\ forms and/or their derivatives (which gives a tempered
distribution). Obviously we will never consider products of delta forms
localized on the same variables.

Sometimes we will also consider more general dependence as $f(d\theta)$ with
$f$ a formal power series in the $d\theta$ variables. The integral over
$d^{n}x$ is the Lebesgue integral, the integrals over $d^{m}\theta$ and
$d^{n}\left(  dx\right)  $ are the Berezin integrals and the integral over
$d^{m}\left(  d\theta\right)  $ is a formal operation, denoted again with
$\int_{\mathbb{R}^{m}},$ with many (but not all) of the usual rules of Dirac's
deltas and of ordinary integration in $\mathbb{R}^{m}$.

The integration with respect to the $d^{m}\left(  d\theta\right)  $ "volume
form" must be interpreted in a way consistent with the crucial property
$\delta(d\theta^{\alpha})\delta(d\theta^{\beta})=-\delta(d\theta^{\beta
})\delta(d\theta^{\alpha}).$ This implies that $d\left[  d\theta\right]  $
must be considered as a form-like object in order to satisfy the natural
property:%
\begin{equation}
\int_{\mathbb{R}^{2}}\delta\left(  d\theta\right)  \delta\left(
d\theta^{\prime}\right)  d\left(  d\theta\right)  d\left(  d\theta^{\prime
}\right)  =1
\end{equation}
In the following we will need to represent $\delta\left(  d\theta\right)  $
and $\delta^{\prime}(d\theta)$ as an integral of this kind. A natural choice
is:%
\begin{subequations}
\begin{align}
\int_{\mathbb{R}^{m}}e^{id\theta\cdot b}d^{m}b  &  =\,\delta^{m}%
(d\theta)\label{rappintegrale}\\
\int_{\mathbb{R}^{m}}b_{1}...b_{m}e^{id\theta\cdot b}d^{m}b  &  =(-i)^{m}%
\,\left(  \delta^{\prime}(d\theta)\right)  ^{m} \label{rappintegrale1}%
\end{align}
where the products $\delta^{m}(d\theta)$ and $\left(  \delta^{\prime}%
(d\theta)\right)  ^{m}$ ($m$ here denotes the number of factors) are wedge
products ordered as in $d^{m}b.$ In other words this kind of integrals depends
on the choice of an oriented basis. For example, we must have:
\end{subequations}
\begin{equation}
\delta(d\theta)\delta\left(  d\theta^{\prime}\right)  =\int_{\mathbb{R}^{2}%
}e^{i (d\theta b+d\theta^{\prime}b^{\prime})}dbdb^{\prime}=-\int_{\mathbb{R}^{2}%
}e^{i (d\theta b+d\theta^{\prime}b^{\prime})}db^{\prime}db=-\,\delta
(d\theta^{\prime})\delta\left(  d\theta\right)
\end{equation}

\noindent {\it Note:} we emphasise that $\mathcal{F}$ maps $\Omega^{(p|q)}$ in $\Omega^{(n-p|m-q)},$ and
that the spaces $\Omega^{(p|0)}$ and $\Omega^{(p|m)}$ are {\it finite
dimensional} in the sense that as modules over the algebra of superfunctions they
are generated by a finite number of monomial-type super and integral forms.

\section{Integral representation of the Hodge dual}

Although most of the usual theory of differential forms can be extended
without difficulty to superforms, the extension of the Hodge dual
has proved to be problematic. This extension clearly would be very relevant
in the study of supersymmetric theories.

The formalism of the Grassmannian integral transforms can be used in the
search of this generalization.
{We will describe in this first paper a simple formal procedure for defining and computing the super Hodge dual.
The ``dual" variables entering the computations are considered only as auxiliary integration variables that disappear in the final result; 
a more rigorous treatment with all mathematical details will be given in the forthcoming paper \cite{lavoro3}}

We begin with the case of the Hodge dual for a standard basis in the
appropriate exterior modules. The next paragraph will be devoted to some
generalizations.

We start with the simple example of ordinary differential forms in
$\mathbb{R}^{2}$ viewed as functions in $\mathbb{R}^{2|2}$, and we compute a
sort of {\it partial} Fourier transform $\mathcal{T}$ on the
{\it anticommuting variables} only:%
\begin{equation}\label{FOUTRA}
\mathcal{T}(\omega)\left(  x,dx\right)  =\int_{\mathbb{R}^{0|2}}\omega
(x,\eta)e^{i(dx^{1}\eta_{1}+dx^{2}\eta_{2})}[d^{2}\eta]
\end{equation}
Taking $\omega(x,dx)=f_{0}(x)+f_{1}(x)dx^{1}+f_{2}(x)dx^{2}+f_{12}%
(x)dx^{1}dx^{2}$, one obtains:%
\[
\mathcal{T}(\omega)\left(  x,dx\right)  =f_{12}(x)+if_{2}(x)dx^{1}%
-if_{1}(x)dx^{2}+f_{0}(x)dx^{1}dx^{2}\,.%
\]
It is evident that in order to reproduce the usual Hodge dual for the standard
inner product, a normalization factor dependent on the form degree must be introduced.
To be precise, in presence of a metric $g_{ij}$ on $\mathbb{R}^2$, 
the integrand of the Fourier transform in (\ref{FOUTRA})
is obtained from the original differential form $\omega(x, dx^i)$ substituting $dx^i$ with 
the dual variable $dx^i \rightarrow g^{ij} \eta_j$ in order to preserve the transformation 
properties of the differential form. For more details we refer to \cite{lavoro3}. In the 
present work we use only diagonal metrics for which these details are unimportant.

For $\omega$ a $k$-form in $\mathbb{R}^{n}$ we have:%
\begin{equation}
\star\omega=i^{\left(  k^{2}-n^{2}\right)  }\mathcal{T}(\omega)=i^{\left(
k^{2}-n^{2}\right)  }\int_{\mathbb{R}^{0|n}}\omega(x,\eta)e^{idx\cdot\eta
}[d^{n}\eta]
\end{equation}
This factor can be obtained computing the transformation of the monomial
$dx^{1}dx^{2}...dx^{k}.$
Noting that only the higher degree term in the $\eta$ variables is involved,
and that the monomials $dx^{i}\eta_{i}$ are commuting objects, we have:%
\begin{align*}
\mathcal{T}\left(  dx^{1}...dx^{k}\right)   &  =\int_{\mathbb{R}^{0|n}}%
\eta_{1...}\eta_{k}e^{idx\cdot\eta}[d^{n}\eta]=\\
&  =\int_{\mathbb{R}^{0|n}}\eta_{1...}\eta_{k}e^{i\left(  \sum_{i=1}^{k}%
dx^{i}\eta_{i}+\sum_{i=k+1}^{n}dx^{i}\eta_{i}\right)  }[d^{n}\eta]=\\
&  =\int_{\mathbb{R}^{0|n}}\eta_{1...}\eta_{k}e^{i\sum_{i=1}^{k}dx^{i}\eta
_{i}}e^{i\sum_{i=k+1}^{n}dx^{i}\eta_{i}}[d^{n}\eta]=\\
&  =\int_{\mathbb{R}^{0|n}}\eta_{1...}\eta_{k}e^{i\sum_{i=k+1}^{n}dx^{i}%
\eta_{i}}[d^{n}\eta]=\\
&  =\int_{\mathbb{R}^{0|n}}\frac{i^{n-k}}{\left(  n-k\right)  !}\eta
_{1...}\eta_{k}\left(  \sum_{i=k+1}^{n}dx^{i}\eta_{i}\right)  ^{n-k}[d^{n}%
\eta]
\end{align*}
Rearranging the monomials $dx^{i}\eta_{i}$ one obtains:
\[
\left(  \sum_{i=k+1}^{n}dx^{i}\eta_{i}\right)  ^{n-k} =\left(  n-k\right)
!\left(  dx^{k+1}\eta_{k+1})(dx^{k+2}\eta_{k+2})...(dx^{n}\eta_{n}\right)  =
\]
\[
=\left(  n-k\right)  !(-1)^{\frac{1}{2}(n-k)(n-k-1)}\left(  dx^{k+1}%
dx^{k+2}...dx^{n})(\eta_{k+1}\eta_{k+2}...\eta_{n}\right)
\]
Finally we have:%
\[
\mathcal{T}\left(  dx^{1}...dx^{k}\right)  =
\]
\[
=\int_{\mathbb{R}^{0|n}}\frac{i^{n-k}}{\left(  n-k\right)  !}\eta_{1...}%
\eta_{k}\left(  n-k\right)  !(-1)^{\frac{1}{2}(n-k)(n-k-1)}\left(
dx^{k+1}dx^{k+2}...dx^{n})(\eta_{k+1}\eta_{k+2}...\eta_{n}\right)  [d^{n}%
\eta]=
\]
\[
=\int_{\mathbb{R}^{0|n}}i^{n-k}(-1)^{\frac{1}{2}(n-k)(n-k-1)}(-1)^{k(n-k)}%
\left(  dx^{k+1}dx^{k+2}...dx^{n})(\eta_{1...}\eta_{k})(\eta_{k+1}\eta
_{k+2}...\eta_{n}\right)  [d^{n}\eta]=
\]
\[
=i^{\left(  n^{2}-k^{2}\right)  }(dx^{k+1}dx^{k+2}...dx^{n})
\]
The computation above gives immediately:%
\begin{equation}
i^{\left(  k^{2}-n^{2}\right)  }\mathcal{T}\left(  dx^{1}...dx^{k}\right)
=\star\left(  dx^{1}...dx^{k}\right)
\end{equation}
and%
\begin{equation}
\mathcal{T}^{2}\left(  \omega\right)  =i^{\left(  n^{2}-k^{2}\right)
}i^{\left(  k^{2}\right)  }\left(  \omega\right)  =i^{n^{2}}\left(
\omega\right)  \label{quadratodifourier}%
\end{equation}

that confirm the usual formula:%
\begin{equation}
\star\star\omega=i^{(\left(  n-k)^{2}-n^{2}\right)  }i^{\left(  k^{2}%
-n^{2}\right)  }i^{n^{2}}(\omega)=(-1)^{k(k-n)}(\omega)
\end{equation}

We can generalize this procedure to superforms of zero picture (note that the
spaces of zero picture superforms or maximal picture integral forms are all
finite dimensional) where we have two types of differentials, $d\theta$ and
$dx.$ As before, the integral transform must be performed only on the
differentials:%
\begin{equation}
\mathcal{T}(\omega)(x,\theta,dx,d\theta)=\int_{\mathbb{R}^{m|n}}%
\omega(x,\theta,\eta,b)e^{i(dx\cdot\eta+d\theta\cdot b)}[d^{n}\eta d^{m}b]
\label{trasformatapersuperforme}%
\end{equation}

A zero picture $p-$superform $\omega$ is a combination of a {\it finite
number} of monomial elements of the form:%
\begin{equation}
\rho_{\left(  r,l\right)  }\left(  x,\theta,dx,d\theta\right)  =f(x,\theta
)dx^{i_{1}}dx^{i_{2}}...dx^{i_{r}}\left(  d\theta^{1}\right)  ^{l_{1}}\left(
d\theta^{2}\right)  ^{l_{2}}...\left(  d\theta^{s}\right)  ^{l_{s}}%
\end{equation}
of total degree equal to $p=r+l_{1}+l_{2}+...+l_{s}.$ We denote by $l$ the sum
of the $l_{i}.$ We have also $r\leq n.$

The super Hodge dual on the monomials can be defined as:
\begin{equation}
\star\rho_{\left(  r,l\right)  }=\left(  i\right)  ^{r^{2}-n^{2}}\left(
i\right)  ^{l}\mathcal{T(}\rho_{\left(  r,l\right)  })=\left(  i\right)
^{r^{2}-n^{2}}\left(  i\right)  ^{l}\int_{\mathbb{R}^{m|n}}\rho_{\left(
r,l\right)  }(x,\theta,\eta,b)e^{i\left(  dx\cdot\eta+d\theta\cdot b\right)
}[d^{n}\eta d^{m}b] \label{superhodge1}%
\end{equation}
where we denote again by $\eta$ and $b$ the dual variables to $dx$ and
$d\theta$ respectively and the integral over $d^{m}b$ is understood as
explained in the definitions (\ref{rappintegrale}) and (\ref{rappintegrale1}) .

The coefficient $\left(  i\right)  ^{l}$ is introduced in order to avoid
imaginary factors in the duals. However this choice of the coefficient is not
unique and has important consequences on the properties of the double dual.

As a simple example we take in $\mathbb{R}^{2|2}$ the form $\rho_{\left(
1,2\right)  }=dx^{1}d\theta^{1}d\theta^{1}\in\Omega^{(3|0)};$ we have:%
\[
\star\rho_{\left(  1,2\right)  }=\left(  i\right)  ^{-3}\left(  i\right)
^{2}\int_{\mathbb{R}^{2|2}}\eta_{1}\left(  b_{1}\right)  ^{2}e^{i\left(
dx\cdot\eta+d\theta\cdot b\right)  }[d\eta_{1}d\eta_{2}db_{1}db_{2}%
]=dx^{2}\delta^{\left(  2\right)  }(d\theta^{1})\delta(d\theta^{2})\in
\Omega^{(-1|2)}%
\]
where $\delta^{\left(  2\right)  }(d\theta^{1})$ is the second derivative and
we use the natural result (the index $\alpha$ is fixed):%
\begin{equation}
\int_{\mathbb{R}}\left(  b_{\alpha}\right)  ^{k}e^{id\theta^{\alpha}b_{\alpha
}}db_{\alpha}=-i\,\delta^{(k)}(d\theta^{\alpha})
\end{equation}

The $\star$ operator on monomials can be extended by linearity to generic
forms in $\Omega^{(p|0)}:$%

\[
\star:\Omega^{(p|0)}\longrightarrow\Omega^{(n-p|m)}%
\]
Both spaces are {\it finite dimensional} and $\star$ is an isomorphism.

An important example in $\mathbb{R}^{n|m}$ is $1\in\Omega^{(0|0)}$:%
\[
\star1=d^{n}x\delta^{m}(d\theta)\in\Omega^{(n|m)}%
\]

In the case of $\Omega^{(p|m)},$ a $m-$ picture $p-$integral form $\omega$ is
a combination of a {\it finite number} of monomial elements as follows:%
\begin{equation}
\rho_{\left(  r|j\right)  }\left(  x,\theta,dx,d\theta\right)  =f(x,\theta
)dx^{i_{1}}dx^{i_{2}}...dx^{i_{r}}\delta^{\left(  j_{1}\right)  }\left(
d\theta^{1}\right)  \delta^{\left(  j_{2}\right)  }\left(  d\theta^{2}\right)
...\delta^{\left(  j_{m}\right)  }\left(  d\theta^{m}\right)
\end{equation}
where $p=r-\left(  j_{1}+j_{2}+...+j_{m}\right)  .$ We denote by $j$ the sum
of the $j_{i}.$ We have also $r\leq n$.

The Hodge dual is:%
\begin{equation}
\star\rho_{\left(  r|j\right)  }=\left(  i\right)  ^{r^{2}-n^{2}}\left(
i\right)  ^{j}\int_{\mathbb{R}^{m|n}}\rho_{\left(  r|j\right)  }(x,\theta
,\eta,b)e^{i\left(  dx\cdot\eta+d\theta\cdot b\right)  }[d^{n}\eta d^{m}b]
\label{superhodge2}%
\end{equation}
which extends the zero picture case to the maximal picture case in which
{\it all }delta forms (or their derivatives) are present.

As a simple example we take in $\mathbb{R}^{2|2}$ the form $\star\rho_{\left(
1,2\right)  }$ computed in the example above:
\[
\star\rho_{\left(  1,2\right)  }=\rho_{\left(  1|2\right)  }=dx^{2}%
\delta^{\left(  2\right)  }(d\theta^{1})\delta(d\theta^{2})\in\Omega
^{(-1|2)}.
\]

We have:%
\[
\star\rho_{\left(  1|2\right)  }=\left(  i\right)  ^{1^{2}-2^{2}}\left(
i\right)  ^{2}\int_{\mathbb{R}^{2|2}}\eta_{2}\delta^{\left(  2\right)  }%
(b_{1})\delta(b_{2})e^{i\left(  dx\cdot\eta+d\theta\cdot b\right)  }[d\eta
_{1}d\eta_{2}db_{1}db_{2}]=-dx^{1}(d\theta^{1})^{2}=-\rho_{\left(  1,2\right)
}\in\Omega^{(3|0)}%
\]
In this particular case $\star\star=-1.$

The iterated transformation is, in this generalized case (note that the
transformation does not change the number $l)$:%
\begin{equation}
\mathcal{T}^{2}\left(  \rho_{\left(  r,l\right)  }\right)  =i^{n^{2}}%
(-i)^{2l}\rho_{\left(  r,l\right)  } \label{doppiatrasformazionegeneralizzata}%
\end{equation}

The double dual on monomials is then given by:%
\begin{equation}
\star\star\rho_{\left(  r,l\right)  }=\left(  i\right)  ^{\left(  \left(
n-r\right)  ^{2}-n^{2}\right)  }(i)^{l}\left(  i\right)  ^{\left(  r^{2}%
-n^{2}\right)  }(i)^{l}i^{n^{2}}(-i)^{2l}=\left(  -1\right)  ^{r(r-n)}%
\rho_{\left(  r,l\right)  } \label{doppioduale1}%
\end{equation}


This means that if $n$ is odd $\star\star$ is the identity in $\Omega^{(p|0)}%
$, because $\left(  -1\right)  ^{r(n-r)}=1$ for every $r,$ but for $n$ even
this is not true because $\left(  -1\right)  ^{r(n-r)}$ depends on $r$ and not
on $p.$ One can avoid this unpleasant behaviour by changing the coefficient
$\left(  i\right)  ^{l}$ in the definitions (\ref{superhodge1}) and
(\ref{superhodge2}):%
\[
\left(  i\right)  ^{l}\rightarrow\left(  i\right)  ^{\alpha\left(  l\right)  }%
\]
Taking into account the formula (\ref{doppiatrasformazionegeneralizzata}) we
have:%
\[
\star\star\rho_{\left(  r,p-r\right)  }=\left(  i\right)  ^{\left(  \left(
n-r\right)  ^{2}-n^{2}\right)  }(i)^{\alpha\left(  l\right)  }\left(
i\right)  ^{\left(  r^{2}-n^{2}\right)  }(i)^{\alpha\left(  l\right)
}i^{n^{2}}(-i)^{2l}\rho_{\left(  r,l\right)  }=\left(  -1\right)
^{r(r-n)+\alpha(l)+l}\rho_{\left(  r,l\right)  }%
\]
Finally choosing $\alpha(l)=2pl-l^{2}-nl-l$ (with $l=p-r$) we obtain:
\begin{equation}
\star\star\rho_{\left(  r,p-r\right)  }=\left(  -1\right)  ^{r(r-n)+2pl-l^{2}%
-nl}\rho_{\left(  r,p-r\right)  }=(-1)^{p(p-n)}\rho_{\left(  r,p-r\right)  }%
\end{equation}
With this choice we have, in $\Omega^{(p|0)}$:%
\begin{equation}
\star\star=(-1)^{p(p-n)}%
\end{equation}
We have obtained a nice duality but the price is the possible appearance of some
imaginary factor in the duals of monomials with $l\neq0$.

Note that the modules $\Omega^{(p|q)}$ for $0<q<m$ are {\it not }finitely
generated and hence for them the definition of a Hodge dual is more problematic.

\subsection{Hodge duals for (super)manifolds}

The Hodge dual depends on the choice of a bilinear form (that in the usual bosonic
case is a scalar product or a metric) that gives an identification between the
module of one-forms and its dual. The same is true for the partial Fourier
transform. In this paragraph we provide a mild generalization of the integral
transform, allowing for a change of the basis and the dual basis that is
necessary for the applications to supersymmetry and supersymmetric theories.

We start with the trivial example of $\mathbb{R}.$

If we denote by $\left\{  1,dx\right\}  $ the basis of the $0-$forms and $1-$
forms respectively, a metric $g^{-1}$ on $\bigwedge^{1}$ is simply a positive
rescaling $dx\rightarrow g^{11}dx$. As usual, we denote by $g_{11}=\left(
g^{11}\right)  ^{-1},$ the rescaling of vector fields {\it and of the dual
variable} $\eta$ (the double dual of vectors).

For this metric the Hodge dual is:%
\begin{equation}
\star1=\sqrt{g_{11}}dx~~~~\text{ and }~~~~\star dx=\frac{1}{\sqrt{g_{11}}}%
\end{equation}
The one form $\sqrt{g_{11}}dx$ is the volume form of the metric.

We can recover this through a small modification of the integral transform
$\mathcal{T}$ procedure.

We introduce a change of basis in $\bigwedge^{1}:$ $dx\rightarrow dx^{\prime
}=Adx$; this rescaling affects also the dual variable: $\eta\rightarrow
\eta^{\prime}=\frac{1}{A}\eta$. In this new basis we compute the transform
$\mathcal{T}$%
\begin{align}
\star1  &  =(-i)\mathcal{T}\left(  1\right)  =(-i)\int_{\mathbb{R}^{0|1}%
}e^{idx^{\prime}\cdot\eta^{\prime}}[d\eta^{\prime}]=dx^{\prime}\\
\star dx^{\prime}  &  =\mathcal{T}\left(  dx^{\prime}\right)  =\int
_{\mathbb{R}^{0|1}}\eta^{\prime}e^{idx^{\prime}\cdot\eta^{\prime}}%
[d\eta^{\prime}]=1
\end{align}
We have now obtained the Hodge dual for the metric $g_{11}^{\prime}=1$.
Reverting to the old variable we get the Hodge dual for the metric
$g_{11}=A^{2}.$%
\begin{equation}
\star1=Adx~~~~\text{ and }~~~~\star dx=\frac{1}{A}%
\end{equation}

The same procedure can be applied to $\mathbb{R}^{n}$, using instead an
invertible matrix $A$ to produce the change of basis, and the product $A^{t}A$
to represent the metric.

For differential forms on curved manifolds we can also use the Cartan frames
(vielbeins) $dx^{i}e_{i}^{a}(x)=dx^{\prime a}$, where $i$ and $a$ denotes here
respectively the curved and the flat indices, both running from $1$ to $n$.
The Hodge dual is then obtained by the following integral transform on $k-$
forms:%
\begin{equation}
\star\omega=i^{\left(  k^{2}-n^{2}\right)  }\int_{\mathbb{R}^{0|n}}%
\omega(x,\eta^{\prime})e^{idx^{\prime a}\eta_{a}^{\prime}}[d^{n}\eta^{\prime}]
\label{hodgeusuale}%
\end{equation}
Where again $\eta^{\prime}$ is the dual basis of the basis $dx^{\prime}$.

For example, we have:%
\begin{align*}
\star1  &  =d^{n}x^{\prime}=\det(e)d^{n}x\\
\star d^{n}x^{\prime} =1  &  \Rightarrow\star d^{n}x=\det(e)^{-1}%
\end{align*}
This Hodge dual is clearly the one determined by the metric $g$ with
$\delta_{ab}=g_{ij}e_{a}^{i}e_{b}^{j}$ and $\delta^{ab}=g^{ij}e_{i}^{a}%
e_{j}^{b}$ , where $e_{a}^{i}$ is the inverse vielbein and $\delta_{ab}$ the
flat metric.

For a supermanifold we will denote collectively by $Z^{M}=\left(
x^{m},\theta^{\mu}\right)  $ and $dZ^{M}=\left(  dx^{m},d\theta^{\mu}\right)
$ (with $M=(m,\mu)$, $m=1...n,$ $\mu=1...m)$ respectively the coordinates and
the differentials, and by $Y_{A}$ (with $A=(a,\alpha)$, $a=1...n,$
$\alpha=1...m)$ the variables dual to the differentials.

As before we introduce the super vielbeins $E_{M}^{A}(Z)$ and we define
$dZ^{\prime A}=dZ^{M}E_{M}^{A}(Z)$ (with dual basis $Y_{A}^{\prime}$) the
transformed differential.

In matrix form we have:%
\[
E_{M}^{A}(Z)=\left(
\begin{array}
[c]{cc}%
E_{m}^{a}(Z) & E_{m}^{\alpha}(Z)\\
E_{\mu}^{a}(Z) & E_{\mu}^{\alpha}(Z)
\end{array}
\right)
\]

The partial Fourier transform (recall that we transform only the ``differentials")
 is%
\begin{equation}
\mathcal{T}(\omega)=\int_{\mathbb{R}^{m|n}}\omega(Z,Y^{\prime})e^{idZ^{\prime
A}Y_{A}^{\prime}}[dY^{\prime}] \label{superhodgecurvo}%
\end{equation}
and the super Hodge dual is defined as above, inserting also the suitable
normalization factors of the previous section. This procedure gives the Hodge dual for the flat basis. We can compute the Hodge dual in the curved basis writing the duals of the differentials $dZ^{\prime A}$ in terms of the old ones $dZ^{M}$.
We obtain, for example, $\star1=d^{n}x^{\prime}\delta^{m}(d\theta^{\prime
})=\mathrm{Sdet}(E)d^{n}x\delta^{m}(d\theta),$ the integral top form ("volume
form") of the supermanifold.

\subsection{A Simple Example for $\mathcal{M}^{(1|1)}$}

In generic supermanifolds the calculations are very long and often the
abstract formulae are not very illuminating.

We will consider in this paragraph a simple and exhaustive example. We
consider an orientable supermanifold $\mathcal{M}^{(1|1)},$ locally modelled
on $\mathbb{R}^{(1|1)},$ parametrized by a bosonic coordinate $x$ and a
fermionic one $\theta$.

We take a $\mathbb{Z}_{2}-$ ordered (the first element is odd and the second
is even) basis $\left\{  dx,d\theta\right\}  $ of $\Omega^{(1|0)}$ and a non
singular superbilinear form $\Phi$ on $\Omega^{(1|0)}$ represented, in this
basis, by an even invertible supermatrix $\mathbb{B}_{(1,0)}.$ The general
form of the matrix $\mathbb{B}_{(1,0)}$ can be written, with a certain amount
of foresight (we want to keep as simple as possible the form of the matrix $\mathbb{A}$ below):%
\[
\mathbb{B}_{(1,0)}=\left(
\begin{array}
[c]{cc}%
A^{-2} & \left(  AB\right)  ^{-1}\left(  \frac{\beta}{B}-\frac{\alpha}%
{A}\right)  \theta\\
\left(  AB\right)  ^{-1}\left(  \frac{\beta}{B}+\frac{\alpha}{A}\right)
\theta & -B^{-2}%
\end{array}
\right)
\]
where $\alpha,\beta,A\neq0,B\neq0$ are real numbers and $\mathrm{Sdet}%
\mathbb{B}_{(1,0)}=-B^{2}/A^{2}.$ It is always possible to find an
{\it even} non singular supermatrix $\mathbb{A}$ (that gives an even
automorphism of $\Omega^{(1|0)}$ that preserves the $\mathbb{Z}_{2}-$ order)
in such a way that $\mathbb{B}_{(1,0)}$ is transformed in the standard
(normalized and diagonal) form:%
\[
\mathbb{A}^{t}\mathbb{B}_{(1,0)}\mathbb{A}=\left(
\begin{array}
[c]{cc}%
1 & 0\\
0 & -1
\end{array}
\right)
\]
This formula suggests that $\mathbb{A}$ can be viewed as the supervielbein mapping the flat metric to the
curved one.
We have:
\begin{equation}
\mathbb{A}=\left(
\begin{array}
[c]{cc}%
A & \alpha\theta\\
\beta\theta & B
\end{array}
\right)  \text{ with }\mathbb{A}^{-1}=\left(
\begin{array}
[c]{cc}%
A^{-1} & -\left(  AB\right)  ^{-1}\alpha\theta\\
-\left(  AB\right)  ^{-1}\beta\theta & B^{-1}%
\end{array}
\right)  \text{ and }\mathbb{A}^{t}=\left(
\begin{array}
[c]{cc}%
A & \beta\theta\\
-\alpha\theta & B
\end{array}
\right)
\end{equation}
We recall that an even matrix is invertible if and only if the even blocks on
the diagonal are invertible, that the transpose is a duality of period $4,$
and that $\mathrm{Sdet}\mathbb{A}=A/B.$ The new basis of one-forms is: $\left\{
dx^{\prime},d\theta^{\prime}\right\}  =\left\{  dx,d\theta\right\}
\mathbb{A}.$ The corresponding new dual basis of $\left(  \Omega^{(1|0)}%
\right)  ^{\ast}$ will be denoted by
$
\left\{
\begin{array}{c}
 \eta^{\prime} \\  b^{\prime}
\end{array}
\right\}
 =\mathbb{A}^{-1}\left\{
\begin{array}{c}
 \eta \\  b
\end{array}
\right\}
 .$ In addition, the
entries of the matrix could in principle become $x$-dependent (if
$\mathbb{B}_{(1,0)}$ is $x$-dependent). We have:%
\begin{equation}
dx^{\prime}=Adx+\theta\beta d\theta~~~~~~~
\text{ and }~~~~~~~~
d\theta^{\prime}=Bd\theta
-\alpha\theta dx
\end{equation}
The partial transform is:
\begin{equation}
\mathcal{T}(\omega)(x,\theta,dx^{\prime},d\theta^{\prime})=\int\omega
(x,\theta,\eta^{\prime},b^{\prime})e^{i(dx^{\prime}\cdot\eta^{\prime}%
+d\theta^{\prime}\cdot b^{\prime})}[d\eta^{\prime}db^{\prime}]\label{exaA}%
\end{equation}

For example:
\begin{equation}
\star1=(-i)\int e^{i(dx^{\prime}\eta^{\prime}+d\theta^{\prime}b^{\prime}%
)}[d\eta^{\prime}db^{\prime}]=dx^{\prime}\delta(d\theta^{\prime})=\left(
\mathrm{Sdet}\mathbb{A}\right)  dx\delta(d\theta)=\sqrt{\left\vert
\mathrm{Sdet}\mathbb{B}_{(1,0)}^{-1}\right\vert }dx\delta(d\theta)
\label{dualediuno}%
\end{equation}
which is a $\Omega^{(1|1)}$ integral top form (that is a "volume form") for
the supermanifold\footnote{We started with an inverse metric, that is a metric
on the $1-$ forms, and hence the "usual" factor $\sqrt{\left\vert g\right\vert
}$ must be substituted here by $\sqrt{\left\vert g^{-1}\right\vert }.$}.

For the Hodge dual of $dx\delta(d\theta)$ we can compute as follows:%
\begin{equation}
\star dx^{\prime}\delta(d\theta^{\prime})=\int\eta^{\prime}\delta(b^{\prime
})e^{i(dx^{\prime}\eta^{\prime}+d\theta^{\prime}b^{\prime})}[d\eta^{\prime
}db^{\prime}]=1\Longrightarrow\star dx\delta(d\theta)=\left(  \mathrm{Sdet}%
\mathbb{A}\right)  ^{-1} \label{dualeditop}%
\end{equation}
The equations (\ref{dualediuno}) and (\ref{dualeditop}) imply  $\star\star=1$
in $\Omega^{(0|0)}$.

Let us consider now the Hodge duals of the $(1|0)$-forms $dx^{\prime}$ and
$d\theta^{\prime}$ and of the $(0|1)$-forms $\delta^{(1)}\left(
d\theta^{\prime}\right)  dx^{\prime}$ and $\delta\left(  d\theta^{\prime
}\right). $The Hodge dual is computed using the partial Fourier transform
$\mathcal{T}$ as follows:
\begin{align}
\star dx^{\prime}  &  =\star\rho_{\left(  1,0\right)  }=\left(  i\right)
^{1^{2}-1^{2}}\left(  i\right)  ^{0}\mathcal{T}(dx^{\prime})=\int\eta^{\prime
}e^{i(dx^{\prime}\eta^{\prime}+d\theta^{\prime}b^{\prime})}[d\eta^{\prime
}db^{\prime}]=\delta\left(  d\theta^{\prime}\right)  \text{ }\nonumber \\
\star d\theta^{\prime}  &  =\star\rho_{\left(  0,1\right)  }=\left(  i\right)
^{0^{2}-1^{2}}\left(  i\right)  ^{1}\mathcal{T}(d\theta^{\prime})=\int
b^{\prime}e^{i(dx^{\prime}\eta^{\prime}+d\theta^{\prime}b^{\prime})}%
[d\eta^{\prime}db^{\prime}]=dx^{\prime}\delta^{(1)}\left(  d\theta^{\prime
}\right)\nonumber  \\
\star\delta\left(  d\theta^{\prime}\right)   &  =\star\rho_{\left(
0|1\right)  }=\left(  i\right)  ^{0^{2}-1^{2}}\left(  i\right)  ^{0}%
\mathcal{T(}\delta\left(  d\theta^{\prime}\right)  )=-i\int\delta\left(
b^{\prime}\right)  e^{i(dx^{\prime}\eta^{\prime}+d\theta^{\prime}b^{\prime})%
}[d\eta^{\prime}db^{\prime}]=dx^{\prime}\nonumber \\
\star dx^{\prime}\delta^{(1)}\left(  d\theta^{\prime}\right)   &  =\star
\rho_{\left(  1|1\right)  }=\left(  i\right)  ^{1^{2}-1^{2}}\left(  i\right)
^{1}\mathcal{T}(dx^{\prime}\delta^{(1)}\left(  d\theta^{\prime}\right)
)=i\int\eta^{\prime}\delta^{(1)}\left(  b^{\prime}\right)  e^{i(dx^{\prime}%
\eta^{\prime}+d\theta^{\prime}b^{\prime})}[d\eta^{\prime}db^{\prime}%
]=d\theta^{\prime} \nonumber%
\end{align}
This is the Hodge dual that corresponds to the bilinear form in $\Omega
^{(1|0)}$ given, in the ordered basis $\left\{  dx^{\prime},d\theta^{\prime
}\right\}  ,$ by the matrix $%
\begin{pmatrix}
1 & 0\\
0 & -1
\end{pmatrix}
$.
Note that the $-1$ on the diagonal is due to the choice of the normalization factor $i^{\left(  r^{2}-n^{2}\right)  }(i)^{l}$ in the definition of the Hodge dual. The other choice $i^{\left(  r^{2}-n^{2}\right)  }(i)^{a(l)}$ (discussed in section 3) gives as diagonal form:
$\begin{pmatrix}
1 & 0\\
0 & 1
\end{pmatrix}
$.

In the original variables we get\footnote{We need to compute the delta form
$\delta\left(  d\theta^{\prime}\right)  $ in terms of $\delta\left(
d\theta\right)  $. This can be done using the formal series (where we denote
by $u$ and $v$ bosonic variables) $\delta\left(  u+v\right)  =\sum_{j}\frac
{1}{j!}\delta^{(j)}(u)\left(  v\right)  ^{j}$. If $u=Bd\theta$ and
$v=-\alpha\theta dx$ (that is nilpotent), the infinite formal sum reduces to a
finite number of terms.}:
\begin{equation}
\star dx=\frac{1}{AB}\delta(d\theta)-\frac{1}{B^{2}}(\frac{\alpha}{A}%
+\frac{\beta}{B})\theta dx\delta^{\prime}(d\theta),\text{ }\nonumber
\end{equation}%
\begin{equation}
\star d\theta=-\frac{1}{B^{2}}(\frac{\beta}{B}-\frac{\alpha}{A})\theta\delta
(d\theta)+\frac{A}{B^{3}}dx\delta^{\prime}(d\theta) \label{exaD}\nonumber %
\end{equation}%
\begin{equation}
\star\delta(d\theta)=ABdx+B^{2}(\frac{\alpha}{A}+\frac{\beta}{B})\theta
d\theta,\text{ }\nonumber
\end{equation}%
\begin{equation}
\star dx\delta^{\prime}(d\theta)=B^{2}(\frac{\beta}{B}%
-\frac{\alpha}{A})\theta dx+\frac{B^{3}}{A}d\theta\,. \label{exaD2}%
\end{equation}
This is the Hodge dual that corresponds to the bilinear form in $\Omega
^{(1|0)}$ given, in the ordered basis $\left\{  dx,d\theta\right\}  ,$ by the
matrix $\mathbb{B}_{(1,0)}.$ We have, for $\phi,\psi\in\Omega^{(1|0)},$ the
standard property
\begin{equation}
\phi\wedge\star\psi=\Phi(\phi,\psi)\star1.
\end{equation}
Note that our super Hodge dual is indeed a duality, and it is an even operator
because it respects the $\mathbb{Z}_{2}$ parity. The set of equations
(\ref{exaD})  provides an explicit isomorphism between
$\Omega^{(1|0)}$ and $\Omega^{(0|1)}$ that can be represented (in the choosen
basis) by a two-by-two supermatrix $\mathbb{G}_{(1|0)}$. The present example
can be exported to
$\Omega^{(p|0)}$ and $\Omega^{(1-p|1)}$,
since these modules are generated by two monomial forms and therefore the
derivation is analogous to that just presented.
Nonetheless, it illustrates the construction of the supermatrix $\mathbb{G}%
_{(p|0)}$ that represents the Hodge dual for the module of $(p|0)$ superforms.
In the following we will adopt the above calculations as a model to discuss
also higher dimensional cases and their relations with physical models.

\subsubsection{Supersymmetry}

Before discussing higher dimensional models, we study the compatibility of the
Hodge dual with supersymmetry. This is important since the present formalism
is adapted to construct supersymmetric Lagrangians. Following the explicit
computations of the previous paragraph we will discuss the case of
$\mathbb{R}^{(1|1)}.$

Unfortunately this case is simple with respect to computations, but it is not
at all simple from the mathematical point of view because the naive
interpretation of the supermanifold $\mathbb{R}^{(1|1)}$ we have adopted since
now, that is a space in which there are "points" with commuting and
anticommuting coordinates $\left(  x,\theta\right)  $ is not adequate. The
main reason  is that in the naive interpretation of $\mathbb{R}^{(1|1)}$ there
is only one real coordinate $x$ and only one fermionic coordinate $\theta
\ $so for supersymmetry we are forced to introduce transformations of coordinates that
apparently are not allowed or meaningful.

Note however that the naive and usual interpretation of supermanifolds is
perfectly valid in all our previous discussions.

Let us first review a few formal ingredients for supersymmetry in
$\mathbb{R}^{(1|1)}$. The variations of the coordinates, the super derivative
and the supersymmetry generators are given by
\begin{equation}
\delta_{\epsilon}x=\frac{1}{2}\epsilon\theta\,,~~~~\delta_{\epsilon}%
\theta=\epsilon\,,~~~~~D=\partial_{\theta}-\frac{1}{2}\theta\partial
_{x}\,,~~~~Q=\partial_{\theta}+\frac{1}{2}\theta\partial_{x}\,.\label{susA}%
\end{equation}
with the algebra
\begin{equation}
\{D,D\}=- \partial_{x}\,,~~~~~\{Q,Q\}= \partial_{x}%
\,,~~~~~\{Q,D\}=0\,.\label{susB}%
\end{equation}
where $\epsilon$ is the \textquotedblleft infinitesimal\textquotedblright%
\ constant Grassmannian supersymmetry parameter.  If, as usual, we want to
consider $\delta_{\epsilon}\theta=\epsilon\,$as a translation in the (unique)
fermionic direction $\theta$ we must conclude that $\epsilon\theta=0.$ So, if we want
to give the geometrical meaning of a "translation" to $\delta_{\epsilon
}x=\frac{1}{2}\epsilon\theta$ we must introduce an auxiliary Grassmann algebra
with at least {\it two} nilpotents generators $\epsilon_{1}$ and
$\epsilon_{2}.$ In this way $\epsilon\ $and $\theta$ are both interpreted as
linear combinations of $\epsilon_{1}$ and $\epsilon_{2},$ and hence $\epsilon$
and $\theta$ are as usual fermionic and nilpotents, and $\epsilon\theta$ is
not a real number but it is bosonic and different from zero.

This procedure can be formalized rigorously defining the supermanifolds of the
type we are considering as super ringed spaces. In this theory the so called "functor
of points" provides a description of the naive "local coordinates" $\left(
x^{i},\theta^{\alpha}\right)  $ as even and odd sections of the sheaves of the
graded rings entering into the definitions. It is not necessary here to give
the details of these constructions and we refer to \cite{vara} for the general theory
and to \cite{Catenacci} for simple examples.

The vector $\epsilon Q$ is an even vector (both $\epsilon$ and $Q$ are odd quantities) and generates
the supersymmetry transformations on the form fields via the usual Lie derivative
\begin{equation}\label{susC}
\delta_\epsilon \omega = {\cal L}_{\epsilon Q} \omega =
(\iota_{\epsilon Q} d + d \iota_{\epsilon Q}) \omega
\end{equation}
for any form $\omega$. We study the compatibility of the supersymmetry with the Hodge dual directly on the
$(1|0)$-forms and on $(0|1)$-forms.
We have
\begin{equation}\label{susD}
\delta_\epsilon (\star dx) = \delta_\epsilon \left( \frac{1}{AB}\delta(d\theta)-
\frac{1}{B^{2}} \left( \frac{\alpha}{A} + \frac{\beta}{B} \right)\theta dx\delta
^{\prime}(d\theta)\right)
\end{equation}
$$
=\left(  -
\frac{1}{B^{2}} \left( \frac{\alpha}{A} + \frac{\beta}{B} \right) \epsilon dx\delta^{\prime}(d\theta)
- \frac{1}{B^{2}} \left( \frac{\alpha}{A} + \frac{\beta}{B} \right)\theta( - \frac12 \epsilon d\theta) \delta^{\prime}(d\theta) \right)
$$
$$
= -
\frac{1}{B^{2}} \left( \frac{\alpha}{A} + \frac{\beta}{B} \right)
\Big(- \frac12 \epsilon \theta \delta(d\theta) + \epsilon dx \delta^{\prime}(d\theta)\Big)
$$
On the other side we have
\begin{equation}\label{susE}
\star (\delta_\epsilon dx) = \star (- \frac12 \epsilon d\theta) =
 \epsilon \frac{1}{2 B^{2}}\Big(\frac{\beta}{B}-\frac{\alpha}{A}\Big) \theta\delta
(d\theta)+\frac{A}{2 B^{3}} \epsilon dx\delta^{\prime}(d\theta)
\end{equation}
Thus, imposing $\delta_\epsilon (\star dx) = \star (\delta_\epsilon dx)$ we find $A = 2 \beta$ and $\alpha=0$. Therefore, the
matrix $\mathbb{A}$ has a triangular form and the corresponding metric $\mathbb{B}$ is symmetric. This is expected for
rigid supersymmetry and it is interesting to recover here the same result.

We notice that there is also another solution:
$\beta= A =0$. This solution gives a non invertible Hodge operator. Nonetheless, we can proceed to
build actions and supersymmetry representations. This particular solution corresponds to the conventional superspace
construction of supersymmetric actions without making use of the Hodge dual construction.

On the $(0|1)$-forms we find
\begin{equation}\label{susF}
\delta_\epsilon (\star d\theta) = \delta_\epsilon \Big( \frac{A}{B^3} ( dx \delta'(d\theta) - \frac12 \theta \delta(d\theta) \Big) =
\frac{A}{B^3} \delta_\epsilon \Big( \Pi \delta'(d\theta) \Big)  = 0\,.
\end{equation}
and, on the other side, we have
$\star \delta_\epsilon d\theta = 0$.
This implies that the only conditions $ A = 2 \beta$ and $\alpha=0$ are sufficient to guarantee compatibility with supersymmetry.

Let us check also the compatibility conditions for the inverse transformations which are the last two eqs. of (\ref{exaD2}). By using $A = 2 \beta$ and $\alpha=0$,
we observe that
\begin{equation}\label{susG}
\star \delta( d\theta ) = B ( A dx + \beta \theta d\theta) = A B (dx + \frac12 \theta d\theta)  = A B \Pi
\end{equation}
where $\Pi \equiv (dx + \frac12 \theta d\theta)$ is the supersymmetric-invariant $(1|0)$-fundamental form.
Then, we immediately get $\delta_\epsilon \Big(\star \delta(d\theta) \Big)= 0$. On the other hand, we have $\star \delta_\epsilon \delta(d\theta) =0$ since $d\theta$ is also invariant.

Finally, let consider
\begin{equation}\label{susH}
\delta_\epsilon (\star dx \delta^\prime(d\theta) ) = \delta_\epsilon \Big[ AB \Big( \frac{B^2}{A^2} d\theta + \frac12 \theta dx \Big) \Big]
 =  \epsilon \frac{AB}{2}  \Pi\,.
\end{equation}
To be compared with
\begin{equation}\label{susI}
\star \delta_\epsilon( dx \delta^\prime(d\theta)) = \star \left( - \frac12 \epsilon d\theta \delta^\prime(d\theta) \right) =
\star (\frac12 \epsilon \delta(d\theta)) = \epsilon \frac{AB}{2} \Pi\,.
\end{equation}
Again, the conditions $A = 2\beta$ and $\alpha=0$ imply compatibility of the supersymmetry with the
star operation.

We can summarise the complete set of Hodge dualities for the supersymmetric variables
\begin{eqnarray}\label{summA}
\star \Pi &=& \frac{1}{AB} \, \delta(d\theta)\,,
~~~\star d\theta = \frac{A}{B^3} \Pi \, \delta'(d\theta)\,, \nonumber \\
\star \delta(d\theta) &=& AB \, \Pi\,,
~~~~~~~~
\star \Pi \delta'(d\theta) = \frac{B^3}{A} d\theta\,.
\end{eqnarray}

We conclude that the supersymmetric variables $\Pi, d\theta$ and $\delta(d\theta), \Pi \delta'(d\theta)$
are exactly the variables in which the metric is diagonal as discussed in the previous sections.
Therefore, compatibility of Hodge duality with supersymmetry implies the ``diagonal" variables.

\subsubsection{The Lagrangian}

We consider a superfield $\Phi^{(0|0)}$ in the present framework. The general
decomposition is
\begin{equation}
\Phi^{(0|0)}\equiv\Phi(x,\theta)=\varphi(x)+\psi(x)\theta\,.\label{supA}%
\end{equation}
where $\varphi(x)$ and $\psi(x)$ are the component fields and they are bosonic
and fermionic, respectively. The supersymmetry transformations are easily
derived:
\begin{equation}
\delta_{\epsilon}\Phi=\epsilon Q\Phi=\epsilon(-\psi(x)+\frac{1}{2}%
\theta\partial_{x}\varphi)~~~~\longrightarrow~~~\delta_{\epsilon}%
\varphi(x)=-\epsilon\psi(x)\,,~~~~~\delta_{\epsilon}\psi(x)=\frac{1}%
{2}\epsilon\partial_{x}\phi(x)\,.\label{supB}%
\end{equation}
We can also compute the differential of $\Phi$ to get
\begin{equation}
d\Phi=(dx+\frac{1}{2}\theta d\theta)\partial_{x}\Phi+d\theta(\partial_{\theta
}-\frac{1}{2}\theta\partial_{x})\Phi=\Pi\partial_{x}\Phi+d\theta
D\Phi\,.\label{supB}%
\end{equation}
Then we can finally compute its Hodge dual
\begin{equation}
\star d\Phi=\star\Pi\partial_{x}\Phi+\star d\theta D\Phi=\frac{1}{AB}%
\delta(d\theta)\partial_{x}\Phi+\frac{A}{B^{3}}\Pi\,\delta^{\prime}%
(d\theta)D\Phi\,.\label{supC}%
\end{equation}
One way to construct a Lagrangian that gives a supersymmetric action is:
\begin{equation}
\mathcal{L}=d\Phi\wedge\star d\Phi=\Big(\Pi\partial_{x}\Phi+d\theta
D\Phi\Big)\wedge\left(  \frac{1}{AB}\delta(d\theta)\partial_{x}\Phi+\frac
{A}{B^{3}}\Pi\,\delta^{\prime}(d\theta)D\Phi\right)  =\nonumber\label{supD}%
\end{equation}%
\[
=\Big(\frac{1}{AB}(\partial_{x}\Phi)^{2}+\frac{A}{B^{3}}(D\Phi)^{2}%
\Big)\Pi\delta(d\theta)\,.
\]
In the $(1|1)$-dimensional case, $\Pi\wedge\Pi=0$ and the second term
$(D\Phi)^{2}$ vanishes. This Lagrangian\footnote{A more usual Lagrangian for the $(1|1)$ -
dimensional case is instead: $-\partial_{x}\Phi D\Phi dx\delta(d\theta)$ } has a peculiarity: the Berezin integral is one-dimensional and
therefore, the contribution from the Lagrangian must be odd. In the
forthcoming sections we present higher dimensional models.

\section{Supersymmetric Theories}

Having discussed the definition of the star operation and
how it can be used in the space of integral forms, we construct examples of supersymmetric theories.
For that we first define
the irreducible representations (for some of them the role of the star operator is
important) in terms of integral- and super-forms. The way how this is done here is
new and it can be easily generalized to several models in different dimensions.

In particular we define the {\it vector multiplet} in 3d N=1  which requires a constraint in order to describe
the off-shell multiplet.\footnote{We recall that the {\it Wess-Zumino multiplet}
in 3d N=1, represented by a $(0|0)$-form $\Phi^{(0|0)}$
does not require any constraint.} This constraint is known in the literature (see for example \cite{Gates:1983nr}), and
we translate it into the present geometric language.  In the same way, we discuss the {\it multiplet
of a conserved current} in 3d N=1, which has the same d.o.f.'s
of the vector multiplet, but has a different realisation and, when translated in the present formalism,
needs the star operation.

Afterwards, we present {\it chiral and anti-chiral superfields} for 4d N=1 superspace,
again in terms of integral forms. These are written in a way that
can be generalised to other models. In addition, we discuss the case of
the {\it linear superfield}, which again requires the use of the star operator.

Finally, in terms of these superfields we construct the corresponding actions.

\subsection{3d N=1 alias ${\cal M}^{(3|2)}$}

We recall that in 3d N=1, the supermanifold ${\cal M}^{3|2}$ (homeomorphic to
$\mathbb{R}^{3|2}$) is described locally by the coordinates
$(x^m,\theta^\alpha)$, and in terms of these coordinates, we have the following two differential operators
\begin{equation}\label{susy3dA}
D_\alpha = \partial_\alpha - \frac12 (\gamma^m \theta)_\alpha \partial_m\,, ~~~~~~
Q_\alpha = \partial_\alpha + \frac12 (\gamma^m \theta)_\alpha \partial_m\,, ~~~~~~
\end{equation}
a.k.a. superderivative and supersymmetry generator, respectively, with the properties
\footnote{In 3d, we use real and symmetric Dirac matrices $\gamma^m_{\alpha\beta}$. The conjugation matrix is
$\epsilon^{\alpha\beta}$ and a bi-spinor is decomposed as follows $R_{\alpha\beta} = R \epsilon_{\alpha\beta}  + R_m \gamma^m_{\alpha\beta}$ where
$R$ and $R_m$ are
a scalar and a vector, respectively.
In addition, it is easy to show that $\gamma^{mn}_{\alpha\beta} = i \epsilon^{mnp} \gamma_{p \alpha\beta}$.
}
\begin{equation}\label{susy3dB}
\{D_\alpha, D_\beta\} = -  \gamma^m_{\alpha\beta} \partial_m\,, ~~~~~
\{Q_\alpha, Q_\beta\} =  \gamma^m_{\alpha\beta} \partial_m\,, ~~~~~
\{D_\alpha, Q_\beta\} = 0\,.
\end{equation}

Given a $(0|0)$ form $\Phi^{(0|0)}$, to compute its supersymmetry variation we apply the
Lie derivative ${\cal L}_\epsilon$ with
$\epsilon = \epsilon^\alpha Q_\alpha + \epsilon^m \partial_m$ ($\epsilon^m$ are the infinitesimal parameters
of the translations and $\epsilon^\alpha$ are the supersymmetry parameters) and
we have
\begin{equation}
\delta_\epsilon \Phi^{(0|0)} = {\cal L}_\epsilon \Phi^{(0|0)} = \iota_\epsilon d \Phi^{(0|0)} =
\iota_\epsilon \Big(dx^m \partial_m  \Phi^{(0|0)} + d\theta^\alpha \partial_\alpha \Phi^{(0|0)}\Big) =
\end{equation}
$$
= (\epsilon^m + \frac12 \epsilon \gamma^m \theta) \partial_m \Phi^{(0|0)} +
\epsilon^\alpha \partial_\alpha \Phi^{(0|0)} = \epsilon^m \partial_m \Phi^{(0|0)} + \epsilon^\alpha Q_\alpha \Phi^{(0|0)}
$$
In the same way, acting on $(p|q)$ forms, we use the usual Cartan formula
${\cal L}_\epsilon = \iota_\epsilon d + d \iota_\epsilon$.

For computing the differential of $\Phi^{(0|0)}$, we can
use a set of invariant $(1|0)$-forms
\begin{equation}
d \Phi^{(0|0)} = dx^m \partial_m \Phi^{(0|0)} + d\theta^\alpha \partial_\alpha \Phi^{(0|0)} =
\end{equation}
$$
=\Big(dx^m + \frac12 \theta \gamma^m d\theta\Big) \partial_m \Phi^{(0|0)} + d\theta^\alpha D_\alpha \Phi^{(0|0)}
\equiv
\Pi^m \partial_m \Phi^{(0|0)} + \Pi^\alpha D_\alpha \Phi^{(0|0)}
$$
with the property $\delta_\epsilon \Pi^m = \delta_\epsilon \Pi^\alpha =0$. This is relevant for having
$\delta_\epsilon d \Phi^{(0|0)} = d \delta_\epsilon \Phi^{(0|0)}$.

The top form is represented by the current
\begin{equation}\label{top3d}
\omega^{(3|2)} = \epsilon_{mnp} \Pi^m\wedge \Pi^n \wedge \Pi^p \wedge \epsilon^{\alpha\beta} \delta(d\theta^\alpha) \wedge \delta(d\theta^\beta)\,,
\end{equation}
which has the properties:
\begin{equation}\label{top3dB}
d \omega^{(3|2)} = 0\,,     ~~~~~
{\cal L}_{\epsilon} \omega^{(3|2)} =0\,.
\end{equation}

According to the previous sections, we can compute the Hodge dual for the supermanifold ${\cal M}^{3|2}$ with a given
supermetric.
We recall that if we define $A=g\left(  \frac{\partial}{\partial x^{i}},\frac{\partial}{\partial
x^{j}}\right)  $ to be a (pseudo)riemannian metric$\ $and { }$B=\gamma
(\frac{\partial}{\partial\theta^{\alpha}},\frac{\partial}{\partial
\theta^{\beta}})$ to be a symplectic form, the even matrix $\mathbb{G}=%
\begin{pmatrix}
A & 0\\
0 & B
\end{pmatrix}
$ is a supermetric in $\mathbb{R}^{n|m}$ (with obviously $m$ even). $A$ and
$B$ are, respectively, $n\times n$ and  $m\times m$ invertible matrices
with real entries and $\det A\neq0$, $\det B\neq0.$
We have to compute the integral transform, and then we must impose compatibility with supersymmetry.
By simple computations (see also \cite{lavoro3}) we obtain (the wedge symbol is omitted)
\begin{align}
\star1 &  =\sqrt{\left\vert \frac{\mathrm{det}(A)}{\mathrm{det}(B)}\right\vert
}\epsilon_{mnp}dx^{m}dx^{n}dx^{p}\delta^{2}(d\theta)\,,~~~~~ &  &
\in~~~\Omega^{(3|2)}\nonumber\label{sus3A}\\
\star dx^{m}= &  \sqrt{\left\vert \frac{\mathrm{det}B}{\mathrm{det}%
A}\right\vert }A^{mn}\epsilon_{npq}dx^{p}dx^{q}\delta^{2}(d\theta)\,,~~~~~ &
&  \in~~~\Omega^{(2|2)}\nonumber\\
\star d\theta^{\alpha}= &  \sqrt{\left\vert \frac{\mathrm{det}B}%
{\mathrm{det}A}\right\vert }B^{\alpha\beta}\epsilon_{mnp}dx^{m}dx^{n}%
dx^{p}\iota_{\beta}\delta^{2}(d\theta)~~~~~ &  &  \in~~~\Omega^{(2|2)}%
\,,\nonumber\\
\star dx^{m}dx^{n}= &  \sqrt{\left\vert \frac{\mathrm{det}B}{\mathrm{det}%
A}\right\vert }A^{mp}A^{nq}\epsilon_{pqr}dx^{r}\delta^{2}(d\theta)~~~~~ &  &
\in~~~\Omega^{(1|2)}\,,\nonumber\\
\star dx^{m}d\theta^{\alpha}= &  \sqrt{\left\vert \frac{\mathrm{det}%
B}{\mathrm{det}A}\right\vert }A^{mp}B^{\alpha\beta}\epsilon_{pqr}dx^{q}%
dx^{r}\iota_{\beta}\delta^{2}(d\theta)~~~~~ &  &  \in~~~\Omega^{(1|2)}%
\,,\nonumber\\
\star d\theta^{\alpha}d\theta^{\beta}=\sqrt{\left\vert \frac{\mathrm{det}%
B}{\mathrm{det}A}\right\vert } &  B^{\alpha\gamma}B^{\beta\delta}%
\epsilon_{pqr}dx^{p}dx^{q}dx^{r}\iota_{\gamma}\iota_{\delta}\delta^{2}%
(d\theta)~~~~~ &  &  \in~~~\Omega^{(1|2)}\,,
\end{align}
where $A^{mn}$ and $B^{\alpha\beta}$ are the components of the inverse
matrices of $A$ and $B$ introduced above.

If, in addition to supersymmetry, we
also impose Lorentz covariance, then $A^{mn}=A_{0}\eta^{mn}$ and
$B^{\alpha\beta}=B_{0}\epsilon^{\alpha\beta}$. Notice that in order to respect
the correct scaling behaviour, assuming that $\theta$ scales with half of the
dimension of $x$'s, $A_{0}$ has a additional power in scale dimensions w.r.t.
$B_{0}$. The quantities $A_{0}$ and $B_{0}$ are constant.
defined here respects the involutive property $\star^2 = 1$.

\noindent {\it Scalar Superfield}

Let us consider now the simplest superfield, i.e.  {\it the scalar superfield}, for the $N=1$ case. This is a $(0|0)$ form
\begin{equation}\label{3d-A}
\Phi^{(0|0)} = A(x) + \theta^\alpha \psi_\alpha(x) + \frac{\theta^2}{2} F(x) \equiv \Phi~~~ \in ~~~\Omega^{(0|0)}\,,
\end{equation}
containing 2 bosonic degrees of
freedom $A,F$ and 2 fermionic ones $\psi_\alpha$. It forms an irreducible representation of the N=1 supersymmetry
algebra and the supersymmetry transformations can be computed by
$\delta_\epsilon \Phi = {\cal L}_\epsilon \Phi = \epsilon^\alpha Q_\alpha \Phi$.

Then, we have
\begin{equation}\label{3dA}
d \Phi = dx^m \partial_m \Phi + d\theta^\alpha \partial_\alpha \Phi = \Pi^m \partial_m \Phi + d\theta^\alpha D_\alpha \Phi~~~
\in ~~~ \Omega^{(1|0)} \,,
\end{equation}
and, in terms of these variables, it is easy to compute the Hodge dual
\begin{equation}\label{3dB}
\star d \Phi = (\star \Pi^m) \partial_m \Phi + (\star d\theta^\alpha) D_\alpha \Phi =
\end{equation}
$$
=A^m_{~n} \Big(\epsilon^n_{pq} \Pi^p \Pi^q \delta^2(d\theta)\Big) \partial_m \Phi  +
B^\alpha_{~\beta} \Big(\Pi^3 \iota^\beta \delta^2(d\theta)\Big) D_\alpha \Phi
~~~\in ~~~ \Omega^{(2|2)} \,,
$$
where $\Pi^3 \equiv \epsilon_{mnp} \Pi^m\wedge \Pi^n \wedge \Pi^p$ and
$\delta^2(d\theta) \equiv \epsilon^{\alpha\beta} \delta(d\theta^\alpha) \delta(d\theta^\beta)$.

The Lagrangian is
\begin{equation}\label{3dB}
{\cal L}_{3d~WZ} = d\Phi\wedge \star d\Phi =
\Big(  \Pi^m \partial_m \Phi + d\theta^\alpha D_\alpha \Phi\Big) \wedge \Big(  (\star \Pi^m) \partial_m \Phi + (\star d\theta^\alpha) D_\alpha \Phi \Big)=
\end{equation}
$$
= \Big(A^{mn} \partial_m\Phi \partial_n \Phi + B^{\alpha\beta} D_\alpha \Phi D_\beta \Phi\Big)\Pi^3 \delta^{2}(d\theta) ~~~~\in ~~~~\Omega^{(3|2)}\,.
$$
As it can be noticed, the expression for ${\cal L}_{3d~WZ}$ represents the generalisation of the usual bosonic expression.
The first term is the usual expression with the bosonic partial derivatives, the second term is a new term, which implements
correctly the fermionic part. To compute the action, we have to integrate  ${\cal L}_{3d~WZ}$  over the supermanifold ${\cal M}^{(3|2)}$ and this gives
\begin{equation}\label{3dC}
S_{3d~WZ} = \int_{{\cal M}^{(3|2)}} \Big(A^{mn} \partial_m\Phi \partial_n \Phi + B^{\alpha\beta} D_\alpha \Phi D_\beta \Phi\Big)\Pi^3 \delta^{2}(d\theta) =
\end{equation}
$$
= \int_{(x,\theta)} \Big(A^{mn} \partial_m\Phi \partial_n \Phi + B^{\alpha\beta} D_\alpha \Phi D_\beta \Phi\Big) \,.
$$
Therefore, we must expand the expression
in the bracket in terms of $\theta$ up to second order. Notice that the functions $A^{mn}$ and $B^{\alpha\beta}$
are superfields. Thus, we must expand them as well.

First we notice that $A^{mn}(x,\theta) = A^{mn}_0(x) + A^{mn}_1(x) \theta^2\,,$ and in the same way
$B^{\alpha\beta}(x,\theta) = B^{\alpha\beta}_0(x) + B^{\alpha\beta}_1(x) \theta^2\,,$ where the coefficients are functions of $x$ only.
If we impose the rigid supersymmetry, the coefficients $A_0$ and $B_0$ are constant, while $A_1$ and $B_1$ are
zero. Then, the second term reproduces the correct WZ action. The first term, on the other hand, is a
supersymmetric higher derivative contribution. It is easy to check its invariance under supersymmetry. The equations
of motion are affected, without spoiling the stability of the path integral. A mass term can be easily added.
Explicitly, we have
\begin{equation}\label{3dD}
S_{3d~WZ} = \int d^3x \Big[
B_0\Big( \frac12 (\partial A)^2 + \psi \!\not\!\partial \psi + \frac12 F^2 \Big)
+ A_0 \Big( \partial_m A \partial^m F + \psi \partial^2 \psi\Big) \Big]
\end{equation}
where the $A_0$ parameter is dimensionful to respect the total dimension of the action. Thus, in 3d the theory is still
renormalizable, even with these higher derivative terms.
\vskip .5cm
\noindent {\it Vector Superfield}

The next representation is the {\it vector superfield} and we start from a superform $A^{(1|0)}$. Then, we construct
its field strength $F^{(2|0)} = d A^{(1|0)}$, invariant under the supergauge transformation $A^{(1|0)} \rightarrow
A^{(1|0)} + d \Lambda^{(0|0)}$ where $\Lambda^{(0|0)}$ is a superfield.
However, the number of component fields of $A^{(1|0)}$ exceeds
the number of physical degrees of freedom for a vector field (and its superpartner) and therefore we must impose
a constraint to reduce them. For that, we observe that the field strength naturally satisfies  the Bianchi identities
$$
d F^{(2|0)} = 0
$$
and with an additional constraint on the field strength one can find the irreducible representation (see \cite{Gates:1983nr}).
We impose $F_{\alpha\beta} =0$, namely the spinorial components are set to zero. To traslate it into
a more geometrical setting we consider the contraction of $\omega^{(3|2)}$ along two spinorial directions with
tangent vector
$\lambda = \lambda^\alpha D_\alpha$, namely
$$\iota^2_\lambda \omega^{(3|2)} = \lambda^\alpha \lambda^\beta \epsilon_{mnp} \Pi^m \Pi^n \Pi^p
\iota_{\alpha} \iota_{\beta} \delta^2(d\theta)\,,
$$ (which becomes a $(1|2)$ integral form) and
we can set the constraint as
\begin{equation}\label{top3dC}
\iota^2_\lambda \omega^{(3|2)}\wedge F^{(2|0)} \propto (\lambda^\alpha \lambda^\beta F_{\alpha\beta}) \omega^{(3|2)} = 0
\end{equation}
which implies the conventional constraint.
Having imposed the constraint, together with the Bianchi identities, we get
\begin{equation}\label{top3dD}
F^{(2|0)} = F_{mn} \Pi^m\wedge \Pi^n + (W \gamma^m)_\alpha \Pi_m\wedge d\theta^\alpha\,.
\end{equation}
where $F_{mn} = (\gamma_{mn})^\alpha_{~\beta} D_\alpha W^\beta$ and $W^\alpha$ is the superfield known as
{\it gluino field strength}.   It satisfies the additional constraint $D_\alpha W^\alpha =0$ which follows from the
Bianchi identities.

Now, we can compute the Hodge dual of $F^{(2|0)}$ to get
\begin{equation}\label{top3dE}
(\star F)^{(1|2)} = F_{mn} \star ( \Pi^m\wedge \Pi^n )+ (W \gamma^m)_\alpha \star (\Pi_m\wedge d\theta^\alpha)=
\end{equation}
$$
= F_{mn} \epsilon^{mn}_{~~~p} \Pi^p \wedge \delta^2(d\theta) + (W \gamma^m)_\alpha \epsilon^m_{~~np} \Pi^n\wedge \Pi^p \wedge\iota^\alpha \delta^2(d\theta)
$$
and therefore we can build an integral top form as usual
\begin{equation}\label{top3dF}
\star F \wedge F = \Big(
A^{mp} A^{nq} \, F_{mn} F_{pq} + A^{mn} B^{\alpha\beta} (W \gamma_m)_\alpha (\gamma_n W)_\beta\Big) \omega^{(3|2)} =
\end{equation}
$$
=\Big( A_0^2 D_\alpha W^\beta D^\alpha W_\beta + A_0 B_0 W_\alpha W^\alpha \Big) \omega^{(3|2)}
$$
Finally, we can compute the action
\begin{equation}\label{top3dG}
S_{3d YM} = \int_{(x,\theta)} \Big( A_0^2 D_\alpha W^\beta D^\alpha W_\beta + A_0 B_0 W_\alpha W^\alpha \Big)
\end{equation}
where $A_0$ and $B_0$ are constant parameters to be related to coupling constants. Notice that the second term is the
correct  abelian SYM 3d Lagrangian (this can be easily verified by expanding the superfield $W^\alpha$ in components and using the
constraint $D_\alpha W^\alpha =0$ to reduce the number of independent components). That term is rescaled with the parameter $A_0 B_0$ which can be used to normalise correctly the kinetic term. The second term however is a novelty since it gives a higher derivative term (scaled with $A_0^2$). As we already noticed the parameters $A_0$ and $B_0$ have different mass dimensions providing the correct scaling behaviour of the action.

In terms of the present ingredients, we can build a new term as follows. Considering the vector superfield $A^{(1|0)}$,
(subject to the constraints (\ref{top3dC})), and computing its Hodge dual we get
\begin{equation}\label{top3dH}
(\star A)^{(2|2)} = A_m \epsilon^{m}_{~~np} \Pi^n\wedge \Pi^p \delta^2(d\theta) + A_\alpha \Pi^3 \iota^\alpha \delta^2(d\theta)\,.
\end{equation}
With that we can construct the following integral form
\begin{equation}\label{top3dI}
\star A\wedge A = \Big( A_0 \eta^{mn} A_m A_n + B_0 \epsilon^{\alpha\beta} A_\alpha A_\beta\Big) \omega^{(3|2)}
\end{equation}
By using the gauge symmetry, we can set $A^{(1|0)}$ into the form $A^{(1|0)} = A_m \Pi^m + A_\alpha d\theta^\alpha$,
where $A_\alpha = (\gamma^m \theta)_\alpha a_m(x) + \psi_\alpha(x) \theta^2/2$ and
$A_m = a_m + (\psi(x) \gamma_m \theta) + \epsilon_m^{~~np} f_{np}(x)$ where $a_m(x), \psi_\alpha(x)$ and $f_{mn}(x)$
are the gauge field, the gluino and the field strength, respectively.
It can be  shown that
\begin{equation}\label{top3dL}
S_{3d CS} = \int_{(x, \theta)} \Big( A_0 \eta^{mn} A_m A_n + B_0 \epsilon^{\alpha\beta} A_\alpha A_\beta\Big) \propto A_0
\int d^3x \Big( \epsilon^{mnp} a_m \partial_n a_p + \epsilon^{\alpha\beta} \psi_\alpha \psi_\beta\Big)
\end{equation}
by expanding $A_\alpha, A_m$ in components. The result coincides with the super Chern-Simons action in 3d.

\vskip .5cm
\noindent {\it Current Superfield}

The third example we consider is the {\it conserved current} superfield $J^{(1|0)}$. The current superfield contains a conserved
current and a spinor (notice that a conserved current in 3d has two independent degrees of freedom which match those
of a spinor in 3d).

Again, we need to impose a
constraint in order to reduce the amount of independent component fields of the superfield $J^{(1|0)}$
and for that we mimic what is done in the case of pure bosonic manifolds $ d \star J \propto \partial^m J_m {\rm Vol}$ (where
${\rm Vol}$ is the top form of the manifold). For a supermanifold,
we consider again  the $(1|0)$-form $J^{(1|0)}  = J_m \Pi^m + J_\alpha d\theta^\alpha$ and we compute its
Hodge dual
\begin{equation}\label{currA}
\star J = J_m \epsilon^{m}_{~~np} \Pi^n \wedge \Pi^p \delta^2(d\theta) + J_\alpha \Pi^3 \iota^\alpha \delta^2(d\theta)
\end{equation}
which turns out to be a $(2|2)$-integral form. Then we can compute its differential to get an expression
proportional to the top integral form $\Omega^{(3|2)}$
\begin{equation}\label{currB}
d \star J \propto \Big(
A_0 \eta^{mn} \partial_m  J_n + B_0
\epsilon^{\alpha\beta}  D_\alpha J_\beta) \Big) \Omega^{(3|2)} = 0
\end{equation}
In the present case, the role of the star operator is fundamental to obtain the divergence of the superfield and
to impose the conservation of the $(1|0)$ superfield. Using the usual relation between the super derivatives and
the partial derivative $\partial_m$: $\{ D_\alpha, D_\beta\} = - \gamma^m_{\alpha\beta} \partial_m$, we can express the first
term as $- \eta^{mn} \gamma_m^{\alpha\beta} D_\alpha D_\beta J_n$ and thus we have
\begin{equation}\label{currC}
 \Big(
 - \frac12 A_0 \eta^{mn} \gamma^{\alpha\beta}_m D_\alpha D_\beta  J_n + B_0
\epsilon^{\alpha\beta}  D_\alpha J_\beta) \Big) =
 \end{equation}
$$=D_\alpha \Big( - \frac12 A_0 \gamma^{\alpha\beta}_m D_\beta J^m + B_0 \epsilon^{\alpha\beta} J_\alpha\Big) = D_\alpha \tilde J^\alpha =0\,, $$
implying that, once the superfield $J^\alpha$ is redefined as
$\tilde J^\alpha = J^\alpha - \frac{A_0}{2 B_0}  \gamma_m^{\alpha\beta} D_\beta J^m$, the constraints are the same as in the usual
framework. Therefore, the structure of the current superfield is exactly as in the usual case.

\subsection{4d N=1 alias ${\cal M}^{(4|4)}$}

Let us recall some basic elements of supersymmetric representations in 4d.
We consider a supermanifold locally homeomorphic to $\mathbb{R}^{(4|4)}$,
parametrised by $(x^m, \theta^\alpha, \bar\theta^{\dot \alpha})$.
We define the following differential operators
\begin{equation}\label{sus4A}
D_\alpha = \partial_\alpha - \frac12 \bar\theta^{\dot\beta} \partial_{\alpha\dot\beta}\,, ~~~~~~
\bar D_{\dot \alpha} = \partial_{\dot\alpha} - \frac12 \theta^{\beta} \partial_{\dot\alpha \beta}\,, ~~~~~~
\end{equation}
$$
Q_\alpha = \partial_\alpha + \frac12 \bar\theta^{\dot\beta} \partial_{\alpha\dot\beta}\,, ~~~~~~
\bar Q_{\dot \alpha} = \partial_{\dot\alpha} + \frac12 \theta^{\beta} \partial_{\dot\alpha \beta}\,, ~~~~~~
$$
with the algebra
\begin{equation}\label{sus4B}
\{D_\alpha, D_\beta\} = 0\,, ~~~~
\{D_\alpha, \bar D_{\dot \beta}\} = - \partial_{\alpha\dot\beta}\,, ~~~~
\{Q_\alpha, Q_\beta\} = 0\,, ~~~~
\end{equation}
$$
\{Q_\alpha, \bar Q_{\dot \beta}\} = \partial_{\alpha\dot\beta}\,, ~~~~
\{D_\alpha, \bar Q_{\dot \alpha}\} =0\,, ~~~~
\{\bar D_{\dot \alpha}, Q_{\alpha}\} =0\,.
$$
with all other possible anticommutation relations equal to zero.  The partial derivative is
$\partial_{\alpha\dot\alpha} = i \sigma^m_{\alpha\dot\alpha} \partial_m$ where $\sigma^m_{\alpha\dot\alpha}$ are the Pauli
matrices $\{\sigma^m, \sigma^n\} = 2 \eta^{mn} \mathbb{I}$.
The main property is $\partial_{\alpha\dot\beta} \partial^{\dot\beta \beta} =  \delta_\alpha^{~\beta}\partial^2$.

A superfield $\Phi$ is a
function of these coordinates. It can be expanded into polynomials of fermionic
coordinates and the coefficients are called the ``component fields". In the same way,
a $(1|0)$-superform $\omega^{(1|0)}$ can be expanded in fundamental 1-superforms
$(dx^m, d\theta^\alpha, d\bar\theta^{\dot \alpha})$ as follows
\begin{equation}\label{SFa}
\omega^{(1|0)} =
dx^m \omega_m(x^m, \theta^\alpha, \bar\theta^{\dot \alpha}) +
d\theta^\alpha \omega_\alpha(x^m, \theta^\alpha, \bar\theta^{\dot \alpha}) +
d\bar\theta^\alpha \omega_{\dot\alpha}(x^m, \theta^\alpha, \bar\theta^{\dot \alpha})=
\end{equation}
$$
=
\Pi^m \omega'_m(x^m, \theta^\alpha, \bar\theta^{\dot \alpha}) +
d\theta^\alpha \omega'_\alpha(x^m, \theta^\alpha, \bar\theta^{\dot \alpha}) +
d\bar\theta^\alpha \omega'_{\dot\alpha}(x^m, \theta^\alpha, \bar\theta^{\dot \alpha})
$$
where
$(\omega_m, \omega_\alpha, \omega_{\dot \alpha})$ and $(\omega'_m, \omega'_\alpha, \omega'_{\dot \alpha})$ are
the component fields and the two expressions are written in two different bases: $(dx^m, d\theta^\alpha, d\bar\theta^{\dot \alpha})$
and $(\Pi^m, d\theta^\alpha, d\bar\theta^{\dot \alpha})$ with $\Pi^m = dx^m + (\theta \sigma^m d\bar\theta + \bar \theta \bar\sigma^m d\theta)$.
The latter is manifestly supersymmetric and is therefore more suitable to study the irreducible representations.
Notice that $d \Pi^m = 2 d\theta \sigma^m d\bar\theta$. Using the above differential operators, the
supersymmetry transformations are given by
\begin{equation}\label{sus4dA}
\delta_\epsilon x^{\alpha\dot \alpha} = \frac12 \epsilon^\alpha \bar\theta^{\dot\alpha} + \frac12 \bar\epsilon^{\dot \alpha} \theta^\alpha\,, ~~~~~
\delta_\epsilon \theta^\alpha = \epsilon^\alpha \,, ~~~~~
\delta_\epsilon \bar\theta^{\dot\alpha} = \epsilon^{\dot \alpha} \,, ~~~~~
 \end{equation}

Following the previous sections, the Hodge dual (compatible with supersymmetry) is
\begin{eqnarray}\label{sus4dB}
\star 1 &=& \frac{{\rm det} A^{mn}}{{\rm det} B^{\alpha\beta} {\rm det}B^{\dot \alpha\dot \beta}} \Pi^4 \delta^2(d\theta) \delta^2(d\bar\theta)  ~~~~~
\in~~~~~ \Omega^{(4|4)} \nonumber \\
\star \Pi^{m} &=& A^{mn} \epsilon_{npqr} \Pi^p \wedge \Pi^q \wedge \Pi^r \delta^2(d\theta) \delta^2(d\bar\theta)\,,
~~~~~~~~
\in~~~~~ \Omega^{(3|4)}
\nonumber  \\
\star d\theta^\alpha &=& B^{\alpha\beta} \Pi^4 \iota_\beta \delta^2(d\theta) \delta^2(d\bar\theta)\,, ~~~
\in~~~~~ \Omega^{(3|4)} \nonumber  \\
\star d\bar\theta^{\dot\alpha} &=& B^{\dot\alpha\dot\beta} \Pi^4 \iota_{\dot\beta} \delta^2(d\theta) \delta^2(d\bar\theta)\,, ~~~~~
\in~~~~~ \Omega^{(3|4)}
\end{eqnarray}
where $\Pi^4 = \epsilon_{mnpq} \Pi^m \wedge \dots \wedge \Pi^q$
and it turns out that the supersymmetric variables are those in which the Hodge operator is diagonal.
The contractions $ \iota_{\dot \beta} $ and $ \iota_{{\beta}}$ act on the product of delta functions.

\vskip .5cm
\noindent {\it Chiral Superfield}

In 4d with 4 fermionic coordinates $\theta^\alpha, \bar\theta^{\dot \alpha}$,
we can define two chiral currents
\begin{eqnarray}\label{ECA}
J^{(4|2)} = \epsilon_{m_1 \dots m_4} \Pi^{m_1}  \wedge \dots \wedge \Pi^{m_4} \wedge \epsilon_{\alpha\beta}
\delta(d\theta^{\alpha}) \wedge \delta(d\theta^{\beta}) \nonumber \\
\overline{J}^{(4|2)} = \epsilon_{m_1 \dots m_4} \Pi^{m_1}  \wedge \dots \wedge \Pi^{m_4} \wedge \epsilon_{\dot\alpha\dot\beta}
\delta(d\bar\theta^{\dot \alpha}) \wedge \delta(d\bar\theta^{\dot\beta})
\end{eqnarray}
Notice that the differential of $\Pi^{\alpha\dot \alpha}$ is $d \Pi^{\alpha \dot \alpha} = 2 d\theta^\alpha \wedge d\bar\theta^{\dot \alpha}$, and therefore
it is easy to check that both currents are closed: $d J^{(4|2)}= 0$ and $d \bar J^{(4|2)}= 0$.
In terms of these currents we can define a chiral and an anti-chiral field
by setting
\begin{eqnarray}\label{ECB}
J^{(4|2)} \wedge  d \Phi = 0\,, \quad \quad   \overline J^{(4|2)} \wedge d \bar \Phi = 0
\end{eqnarray}
To see this, we compute the differential $d \Phi = d\theta^\alpha D_\alpha \Phi + d\bar\theta^{\dot \alpha} D_{\dot \alpha} \Phi + \Pi^{\alpha\dot\alpha}
\partial_{\alpha\dot\alpha} \Phi$
and we have
\begin{eqnarray}\label{ECC}
&&\epsilon_{m_1 \dots m_4} \Pi^{m_1}  \wedge \dots \wedge \Pi^{m_4} \wedge \epsilon_{\alpha\beta}
\delta(d\theta^{\alpha}) \wedge \delta(d\theta^{\beta}) \wedge ( d\theta^\alpha D_\alpha \Phi + d\bar\theta^{\dot \alpha} D_{\dot \alpha} \Phi + \Pi^{\alpha\dot\alpha}
\partial_{\alpha\dot\alpha} \Phi) \nonumber \\
&& = \epsilon_{m_1 \dots m_4} \Pi^{m_1}  \wedge \dots \wedge \Pi^{m_4} \wedge \epsilon_{\alpha\beta}
\delta(d\theta^{\alpha})  \wedge \delta(d\theta^{\beta}) d\bar\theta^{\dot \alpha} D_{\dot \alpha} \Phi = 0
\end{eqnarray}
from this  $D_{\dot \alpha} \Phi = 0$ follows, since the other terms are automatically set to zero. Analogously, considering the
other equation in (\ref{ECB}) we obtain $D_\alpha \bar\Phi = 0$.

Since there are chiral currents, we can define a chiral integral on the reduced supermanifold ${\cal M}^{(4|2)}$ parametrised
by the coordinates $(x^{\alpha\dot\alpha}, \theta^\alpha)$.\footnote{The relation between these coordinates and the original ones is as usual $x^{\alpha\dot \alpha} \rightarrow x^{\alpha\dot \alpha} + \theta^\alpha \bar \theta^{\dot\alpha}$ for chiral  and
$x^{\alpha\dot \alpha} \rightarrow x^{\alpha\dot \alpha} - \theta^\alpha \bar \theta^{\dot\alpha}$ for antichiral supermanifold.}
The above conditions (\ref{ECB}) are needed to
define a chiral integral invariant under variations
\begin{equation}\label{ECD}
\delta \int_{{\cal M}^{(4|2)}} \Phi J^{(4|2)} = \int {\cal L}_{X} \Big( \Phi J^{(4|2)} \Big) =
\int_{{\cal M}^{(4|2)}} (\iota_X d + d \iota_X) \Big(\Phi J^{(4|2)} \Big) =
\end{equation}
$$
=\int_{{\cal M}^{(4|2)}} \iota_X d  \Big(\Phi J^{(4|2)} \Big) =
\int_{{\cal M}^{(4|2)}} \iota_X \Big(d \Phi \wedge J^{(4|2)} \Big) = 0
$$
where the conditions (\ref{ECB}) and the closure of $J^{(4|2)}$ are used, and boundary terms are neglected.
Then, we can define the integrals of {\it chiral integral forms}.
Of course, if $\Phi$ is chiral, any function of it is also chiral and therefore we can write a general action for a chiral field as
\begin{equation}\label{ECE}
S_{V} = \int_{{\cal M}^{(4|2)}}  V(\Phi) J^{(4|2)}\,.
\end{equation}
For a chiral supermanifold, we can introduce a chiral Hodge dual operator $\star_C$,
by restricting the Fourier transforms to the differentials $dx^{\alpha\dot \alpha}$ and $d\theta^\alpha$, leaving aside
the differentials $d\bar\theta^{\dot\alpha}$ since they do not enter the chiral superfield and superforms (notice that if $A^{(1|0)} \in \Omega^{(1|0)}$
can be expanded as (\ref{SFa}), the condition $J^{(4|2)} \wedge A^{(1|0)} =0$ implies that the component $A_{\dot \alpha}$ must vanish).

An additional term for a 4d action for a superfield is the usual kinetic term
\begin{equation}\label{kinA}
S_K = \int_{{\cal M}^{(4|4)}} \star (\bar\Phi \Phi)\,,
\end{equation}
Notice that the product $\bar\Phi\Phi$ is not chiral ({\it i.e.} $d (\bar\Phi\Phi)\wedge J^{(4|2)} \neq 0$ and
$d (\bar\Phi\Phi)\wedge \bar J^{(4|2)} \neq 0$) and therefore it must be integrated on the complete supermanifold.
Therefore the Hodge dual is the complete Hodge dual of the manifold.

There is another possibility to build a supersymmetric action starting from chiral superfields:
\begin{equation}\label{kinB}
S_{dK} = \int_{{\cal M}^{(4|4)}} d\bar\Phi \wedge \star d\Phi\,,
\end{equation}
which however produces higher derivative terms in the action. Notice  that if the Hodge dual has $\theta$-dependent terms, the
component expansion of (\ref{kinA}) and (\ref{kinB}) share some terms. Nonetheless the latter has higher derivative terms.

If we use the following parametrisation of the Hodge dual
 for the fundamental 1-forms $d\theta^\alpha, d\bar\theta^{\dot\alpha}, dx^\mu$
\begin{eqnarray}\label{HDa}
&& \star \, d \theta^\alpha =
G^\alpha_{\beta} \, \iota_{{\beta}} \delta^4(d\theta) d^4x +
G^\alpha_{\dot\beta} \, \iota_{{\dot\beta}} \delta^4(d\theta) d^4x +
G^\alpha_{\nu} \,  \delta^4(d\theta) (d^3x)^\nu \nonumber \\
&& \star \, d \bar\theta^{\dot\alpha} =
G^{\dot\alpha}_{\beta} \, \iota_{{\beta}} \delta^4(d\theta) d^4x +
G^{\dot \alpha}_{\dot\beta} \, \iota_{{\dot\beta}} \delta^4(d\theta) d^4x +
G^{\dot \alpha}_{\mu} \, \delta^4(d\theta) (d^3x)^\mu \nonumber \\
&& \star \, d x^{\mu} =
G^\mu_{\beta} \, \iota_{{\beta}} \delta^4(d\theta) d^4x +
G^\mu_{\dot\beta} \, \iota_{{\dot\beta}} \delta^4(d\theta) d^4x +
G^\mu_{\nu} \, \delta^4(d\theta) (d^3x)^\nu
\end{eqnarray}
where $\delta^4(d\theta) d^4x = \epsilon^{\alpha\beta} \delta(d\theta^\alpha) \delta(d \theta^\beta)
\epsilon^{\dot\alpha\dot\beta} \delta(d\bar\theta^{\dot\alpha}) \delta(d \bar\theta^{\dot\beta}) \epsilon_{\mu\nu\rho\sigma} dx^\mu \wedge \dots \wedge dx^\sigma$ and
$(d^3x)^\mu = \epsilon^\mu_{~\nu \rho \sigma} dx^\nu \wedge dx^\rho \wedge dx^\sigma$,
the computation of $d\Phi\wedge \star d\Phi$ proceeds as follows.

Given the chiral superfield $\Phi$, discussed above, we  decompose it into its components
\begin{eqnarray}\label{WZb}
\Phi(y^{\alpha\dot\alpha}, \theta^\alpha) &=& A(y^{\alpha\dot\alpha}) + \psi_\alpha(y^{\alpha\dot\alpha}) \theta^\alpha + F(y^{\alpha\dot\alpha}) \theta^2 \nonumber \\
&=&
\left(A(x) + \partial_{\beta\dot\beta} A(x) \theta^\beta \bar\theta^{\dot\beta} +
\frac{1}{2} \partial^2 A(x) \theta^2 \bar\theta^2 \right) \nonumber \\
&+&\left(\psi_\alpha(x) + \partial_{\beta\dot\beta} \psi_\alpha(x)  \theta^\beta \bar\theta^{\dot\beta}\right) \theta^\alpha +
F(x) \theta^2\,.
\end{eqnarray}
where $y^{\alpha\dot\alpha} = x^{\alpha\dot\alpha} + \theta^\alpha \bar\theta^{\dot\alpha}$ and  we compute
its differential:
\begin{eqnarray}\label{WZc}
(d \overline \Phi)_{(1|0)}&=&
\left(\partial_m \bar A + \partial_{\beta\dot\beta} \partial_m \bar A \theta^\beta \bar\theta^{\dot\beta} +
\frac{1}{2} \partial^2 \partial_m \bar A \theta^2 \bar\theta^2 \right) dx^m \nonumber \\
&+&\left(\partial_m \bar\psi_{\dot\alpha} + \partial_{\beta\dot\beta} \partial_m \bar\psi_{\dot\alpha}  \theta^\beta \bar\theta^{\dot\beta}\right) \bar\theta^{\dot\alpha} dx^m +
\partial_m \bar F \bar\theta^2 dx^m \nonumber \\
&+&\partial_{\beta\dot\beta} \bar A (d\theta^\beta \bar\theta^{\dot\beta} + \theta^\beta d \bar\theta^{\dot\beta}) +
 \partial^2 \bar A (\theta^\alpha d\theta_\alpha \bar\theta^2 + \theta^2 \bar\theta^{\dot\alpha} d \bar\theta_{\dot \alpha})  \nonumber \\
&+&\bar\psi_{\dot\alpha} d\bar\theta^{\dot\alpha} + \partial^{~\dot\alpha}_{\beta} \bar\psi_{\dot\alpha} (\bar\theta^{\dot\gamma} d\bar\theta_{\dot\gamma} \theta^{\beta} + \bar\theta^2 d\theta^{\beta})
+ \bar F \, 2 \bar\theta^{\dot\alpha} d\bar\theta_{\dot\alpha}\,;
\end{eqnarray}
Then we have
\begin{equation}\label{WZd}
{\cal L} = (\partial_{\alpha\dot\alpha}\Phi, \partial_\alpha \Phi, \partial_{\dot\alpha} \Phi)
\left(
\begin{array}{ccc}
G^{\alpha\dot\alpha \beta\dot\beta}  & G^{\alpha\dot\alpha \beta}  & G^{\alpha\dot\alpha \dot\beta}   \\
\bullet & G^{\alpha \beta}  & G^{\alpha  \dot\beta}  \\
\bullet  &\bullet  & G^{\dot\alpha \dot\beta}
\end{array}
\right)
\left(
\begin{array}{c}
\partial_{\beta\dot\beta}\bar\Phi    \\
\partial_{\beta}\bar\Phi   \\
\partial_{\dot\beta}\bar\Phi
\end{array}
\right)
\end{equation}
where $\bullet$ denotes the transposed element of the supermatrix. However, the components of
that super matrix could in principle be proportional to $\theta^2$ or $\bar\theta^2$
such as
 \begin{equation}\label{WZe}
\left(
\begin{array}{ccc}
\epsilon^{\alpha\beta} \epsilon^{\dot\alpha \dot\beta} \theta^2 \bar\theta^2  & \epsilon^{\alpha\beta} \epsilon^{\dot\alpha \dot\beta}   \theta^2 \bar\theta_{\dot\beta}   &
\epsilon^{\alpha\beta} \epsilon^{\dot\alpha \dot\beta} \theta_\beta \bar\theta^2   \\
\bullet &   \epsilon^{\alpha\beta} \bar\theta^2    &  \theta^\alpha \bar\theta^{\dot\beta}    \\
\bullet  & \bullet  & \epsilon^{\dot\alpha\dot\beta} \theta^2
\end{array}
\right)
\end{equation}
 and the corresponding terms in (\ref{WZd}) renormalize the kinetic term in (\ref{kinA}).

\vskip .5cm
\noindent {\it Linear Superfield}

There exists another multiplet which can be defined in terms of an integral form.
The {\it linear multiplet} is defined in terms of the $(0|0)$-superform $\Phi^{(0|0)}$.
We start by considering the total differential $d \Phi^{(0|0)}$, which is
a $(1|0)$ superform.
Then we have the sequence of operations
\begin{eqnarray}\label{newECC}
\Phi &&\rightarrow d\Phi \in \Omega^{(1|0)} \nonumber \\
&&\rightarrow J^{(4|2)}\wedge d\Phi \in \Omega^{(5|2)} \nonumber \\
&&\rightarrow \star (J^{(4|2)}\wedge d\Phi) \in \Omega^{(-1|2)} \nonumber \\
&& \rightarrow  \bar J^{(4|2)} \wedge \star (J^{(4|2)}\wedge d\Phi) \in \Omega^{(3|4)} \nonumber \\
&& \rightarrow d( \bar J^{(4|2)} \wedge \star (J^{(4|2)}\wedge d\Phi)) \in \Omega^{(4|4)} \nonumber \\
&& = (\epsilon^{\dot \alpha \dot \beta} \bar D_{\dot \alpha} \bar D_{\dot \beta} \Phi) J^{(4|4)}
\end{eqnarray}
 So, by setting to zero the last expression, one recovers the usual definition, namely $\epsilon^{\alpha\beta} D_\alpha D_\beta \Phi=0$,
 of the linear multiplet. It is interesting that
 we had to pass to negative form degree to define the correct equation. Obviously, the same equation can be constructed also
 for the complex conjugate and one can thus define either the linear real superfield or the linear complex superfield.

 \vskip .5cm
\noindent {\it Vector Superfield}

We consider now another multiplet, {\it the gauge multiplet} which is described by a gauge field (with the
corresponding gauge symmetry), the gaugino and an auxiliary field.
Let us consider the connection $A = A_\alpha d\theta^\alpha + A_{\dot \alpha} d\bar \theta^{\dot \alpha} + A_{\alpha\dot \alpha} \Pi^{\alpha\dot \alpha}$.
We apply the differential
\begin{eqnarray}\label{YMa}
F = d A &=&
\left( D_\alpha A_\beta \right) d\theta^\alpha \wedge d\theta^\beta +
\left( D_{\dot\alpha} A_{\dot\beta} \right) d\bar\theta^{\dot\alpha} \wedge d\bar\theta^{\dot\beta} + \nonumber \\
&+& \left( D_\alpha A_{\dot\beta} + D_{\dot \beta} A_\alpha + A_{\alpha\dot \beta} \right) d\theta^\alpha \wedge d\bar\theta^{\dot\beta} +
\left( D_\alpha A_{\beta\dot \beta} - \partial_{\beta\dot\beta} A_\alpha \right) d\theta^\alpha \wedge \Pi^{\beta\dot \beta} +\nonumber \\
&+&
\left( D_{\dot \alpha} A_{\beta\dot \beta} - \partial_{\beta\dot\beta} A_{\dot \alpha} \right) d\bar \theta^{\dot \alpha} \wedge \Pi^{\beta\dot \beta} +
\left( \partial_{\alpha\dot \alpha} A_{\beta\dot \beta} - \partial_{\beta\dot\beta} A_{\alpha\dot \alpha} \right) \Pi^{\alpha\dot \alpha} \wedge \Pi^{\beta\dot \beta}
\end{eqnarray}

Now, if we impose the conditions
\begin{equation}\label{YMb}
J^{(4|2)} \wedge  F = 0\,, \quad \quad   \overline J^{(4|2)} \wedge F = 0
\end{equation}
we find the constraints $ D_{(\alpha} A_{\beta)} = 0$ and $ D_{(\dot\alpha} A_{\dot\beta)} = 0$.
In this way, we still miss the constraint $\left( D_\alpha A_{\dot\beta} + D_{\dot \beta} A_\alpha + A_{\alpha\dot \beta} \right) =0$.

We can consider however a different approach, taking into account the volume density $J^{(4|4)}$
given by
\begin{equation}\label{YMc}
J^{(4|4)} = \epsilon_{m n r s} \Pi^m \wedge \dots \Pi^s \wedge \delta^2(d\theta)  \delta^2(d\bar\theta)\,,
\end{equation}
which is not chiral. Note that, by using the properties of  the Dirac delta forms, this can be written by substituting
$\Pi^m \rightarrow dx^m$ in the bosonic factor. Now, we can consider the contraction with respect to
a commuting 1-form $d\theta^\alpha$ defined as $\iota_{\alpha}$ (notice that this operator commutes as
$\iota_{\alpha} \iota_{\beta} = \iota_{\beta} \iota_{\alpha}$). Formally,
\begin{equation}\label{YMd}
\iota_{\alpha} = \frac{\partial}{\partial (d\theta^\alpha)}\,, \quad\quad \iota_{{\dot\alpha}} = \frac{\partial}{\partial (d\bar\theta^{\dot\alpha})}
\end{equation}
Then, we can impose the constraints as follows
\begin{eqnarray}\label{YMe}
\left(\iota_{\alpha} \iota_{\beta} J^{(4|4)}\right) \wedge F &=& 0 \,, \nonumber \\
\left(\iota_{\alpha} \iota_{{\dot\beta}} J^{(4|4)} \right)\wedge F &=& 0 \,, \nonumber \\
\left(\iota_{{\dot\alpha}} \iota_{{\dot\beta}} J^{(4|4)} \right) \wedge F &=& 0
\end{eqnarray}
implying $F_{(\alpha\beta)} = F_{\dot\alpha \dot\beta} = F_{\alpha \dot\beta} =0$ which are the usual vector superfield constraints.

There is another way to do it. The vector superfield can also
be constructed out of a spinorial superfield $W^\alpha$ (and its conjugate $\bar W^{\dot \alpha}$). For that we
have the chirality conditions
\begin{eqnarray}\label{YMf}
J^{(4|2)} \wedge d W^\alpha =0\,, \quad \quad
\overline J^{(4|2)} \wedge d \overline W^{\dot\alpha} =0\,, \quad \quad
\end{eqnarray}
This implies the constraints $D_\alpha \overline W_{\dot \beta} =0$ and $\overline D_{\dot \alpha} W_\beta =0$.
The additional constraint $D_\alpha W^\alpha + \overline D^{\dot \alpha} \overline W_{\dot \alpha} =0$ is obtained as follows
\begin{equation}\label{YMg}
\left(\iota_{\alpha} J^{(4|4)}\right) \wedge d W^\alpha +
\left(\iota_{{\dot \alpha}} J^{(4|4)}\right) \wedge d \overline W^{\dot \alpha} =0
\end{equation}
It is easy to see that this indeed produces the correct constraints. The equations for the constraints are very geometrical since
they tell us that the field strengths have non vanishing components only in the bosonic directions.

Imposing the constraints, we can rewrite the field strength $F^{(2|0)}$ as follows
\begin{equation}\label{fsA}
F^{(2|0)} = F_{mn} \Pi^m\wedge \Pi^n +
\bar W^{\dot \alpha} \Pi^m (\gamma_m d\theta)_{\dot\alpha} +
W^{\alpha} \Pi^m (\gamma_m d\bar\theta)_{\alpha}
\end{equation}
and then compute its Hodge dual. We thus obtain
the action
\begin{equation}\label{fsB}
S_{SYM} = \int_{{\cal M}^{(4|4)}} F \wedge \star F = \int_{(x,\theta, \bar\theta)} \Big( A_0^2 F^{mn} F_{mn} +
A_0 B_0 W^\alpha W_\alpha + A_0 \bar B_0 \bar W^{\dot\alpha} W_{\dot\alpha} \Big)\,.
\end{equation}
Here we denote $A_0, B_0$ and $\bar B_0$ as the constant overall normalizations of $A^{mn} = A_0 \eta^{mn},
B^{\alpha\beta} = B_0 \epsilon^{\alpha\beta}$ and $\bar B^{\dot\alpha\dot\beta} = \bar B_0 \epsilon^{\dot \alpha\dot\beta}$.
The second and the third terms reproduce the correct vector supefield action (with the $\theta$-term and the coupling constant as
a combination of the two parameters $B_0$ and $\bar B_0$). The first term, however, is a higher derivative term (with
the dimensionful parameter  $A_0$), and it can be expressed in terms of covariant derivatives of $W^\alpha$ and
$\bar W^{\dot \alpha}$.

\section{Summary}

We summarise in a table the 3d and 4d results
discussed in the previous sections.

\begin{table}[ht]
\caption{Summary of models}
\vskip .2cm
\centering
\begin{tabular}{|c| c| c|}
\hline\hline
Case & 3d & 4d  \\ [0.5ex] 
\hline
~&~&~ \\
Potential &$~~\int_{{\cal M}^{3|2}} \star V(\Phi)~~$&~~ $\int_{{\cal M}^{4|2}_C} \star_C V(\Phi) + {\rm c.c.}~~$  \\
~&~&~ \\
Kinetic term &$~~\int_{{\cal M}^{3|2}} d \Phi \wedge \star d \Phi$ ~~&~~ $\int_{{\cal M}^{4|4}} \star \Phi \bar \Phi~~$  \\
~&~&~ \\
Cosm. Cons &$~~\int_{{\cal M}^{3|2}} \star 1$ ~~&~~ $\int_{{\cal M}^{4|2}_C} \star_C 1 + {\rm c.c.}~~$  \\
~&~&~ \\
Hilbert-Einstein & $~~\int_{{\cal M}^{3|2}} \star R$ ~~&~~ $\int_{{\cal M}^{4|4}} \star 1 ~~$ \\ [1ex]
\hline
\end{tabular}``diagonal"
\label{table:nonlin}
\end{table}
This symbol $\star_C$ denotes the Hodge dual in the chiral supermanifold  ${\cal M}^{(4|2,0)}$ or ${\cal M}^{(4|0,2)}$ (in the
table we used the notation $(4|2)$ for readability). In 4d, the superfield $\Phi$ is chiral according to the previous
section. The integrals, both in $3d$ and in $4d$ are on the entire supermanifold, without taking into account
possible boundary contributions.

\section{Acknowledgements}
We would like to thank L. Andrianopoli, R. D'Auria, M. Trigiante and S. Ferrara for several interesting
discussions on integral forms and supersymmetry.

\vfill
\eject

 \appendix

\noindent{\LARGE \bf Appendices}

 \section{Fourier transform and cohomology}

We will discuss in this appendix A and in the following appendix B some relations between Fourier transforms and cohomology.
Here we limit ourselves to some preliminary observations, leaving more
insights and applications to subsequent publications.

Recall that if $M$ is a bosonic manifold with cotangent bundle ${\Omega
^{\bullet}(}M{),}$ a section\ $\omega$ of ${\Omega^{\bullet}(}M{)}$ is viewed
locally as a function on a supermanifold $\mathcal{M}$ of dimension $n|n$ with
local coordinates $(x^{i},dx^{i}).$ We introduce now new {\it fermionic}
coordinates $\theta_{i}$ and their {\it bosonic} differentials $d\theta
_{i}$ that we will consider as (dual) coordinates $(d\theta_{i},\theta_{i})$
on a supermanifold $\mathcal{M}^{\star}.$ With this notations, if
\ $\omega(x,dx)$ is a differential form, its Fourier image is written locally
(see \ref{fourier4}) as:
\begin{equation}
\mathcal{F}(\omega)\left(  d\theta,\theta\right)  =\int\omega
(x,dx)e^{i(d\theta_{i}x^{i}+\theta_{i}dx^{i})} \label{fourierduale}%
\end{equation}

Here and in the following, in order to shorten the notations, we will often
omit the ``integration measure" and the space on which the integration is performed.

As an example we consider the cohomology of the circle $\mathbb{S}^{1}$
and we will map it into a cohomology of integral forms. We consider
$\mathbb{S}^{1}\subset\mathbb{R}^{2}$ given by $x^{2}+y^{2}=1$ and
$xdx+ydy=0.$ (The nontrivial cohomologies in this example arise from
{\it both} relations). The generators of the $d-$ cohomology are given
locally by:
\begin{equation}
H^{0}(\mathbb{S}^{1})=\left\{  1\right\}  \,,\quad H^{1}(\mathbb{S}%
^{1})=\left\{  xdy-ydx\right\}  \label{sph}%
\end{equation}
We take $\omega=1+xdy-ydx$ and we compute locally its Fourier transform
$\mathcal{F}(\omega)$ by introducing the coordinates $\theta_{i}$ and their
differentials $d\theta_{i}$ to get:
\begin{align}
\mathcal{F}(\omega)\left(  d\theta,\theta\right)   &  =\int
(1+xdy-ydx)e^{i(d\theta_{1}x+d\theta_{2}y+\theta_{1}dx+\theta_{2}%
dy)}=\,\label{sphB}\\
&  =\theta_{1}\theta_{2}\,\delta(d\theta_{1})\delta(d\theta_{2})+\theta
_{1}\,\delta^{\prime}(d\theta_{1})\delta(d\theta_{2})+\theta_{2}%
\,\delta(d\theta_{1})\delta^{\prime}(d\theta_{2}) \label{sphB1}%
\end{align}
The result spans the following cohomology spaces:
\begin{align}
H^{(0|2)}(\mathcal{S}^{1\star})  &  =\left\{  \theta_{1}\theta_{2}%
\,\delta(d\theta_{1})\delta(d\theta_{2})\right\} \\
H^{(-1|2)}(\mathcal{S}^{1\star})  &  =\left\{  \theta_{1}\,\delta^{\prime
}(d\theta_{1})\delta(d\theta_{2})+\theta_{2}\,\delta(d\theta_{1}%
)\delta^{\prime}(d\theta_{2})\right\}
\end{align}
It is easy to check that $\theta_{1}\theta_{2}\,\delta(d\theta_{1}%
)\delta(d\theta_{2})$ and $\theta_{1}\,\delta^{\prime}(d\theta_{1}%
)\delta(d\theta_{2})+\theta_{2}\,\delta(d\theta_{1})\delta^{\prime}%
(d\theta_{2})$ are closed but not exact and belong to the cohomology of the
differential $d$ of the ``dual supermanifold" $\mathcal{S}^{1\star}$. For more details on the
cohomology of superforms and integral forms see \cite{Catenacci:2010cs}. The first
generator $\theta_{1}\theta_{2}\,\delta(d\theta_{1})\delta(d\theta_{2})$
corresponds to a {\it picture changing} operator for the supermanifold
$\mathcal{S}^{1\star}$. We will differ to the appendix C some observations on the picture changing
operators with integral forms.

Let us consider now the representation of the cohomology classes using the
angular variable $\varphi$, its differential $d\varphi,$ and the dual
variables $(d\theta,\theta)$. Then, we have
\begin{equation}
H^{0}(\mathbb{S}^{1})=\left\{  1\right\}  \,,\quad H^{1}(\mathbb{S}%
^{1})=\left\{  d\varphi\right\}  \label{ang}%
\end{equation}
and we set $\omega=\alpha+\beta d\varphi.$ We perform the Fourier transform as
follows
\[
\mathcal{F}(\omega)=\int\omega(\varphi, d\varphi) e^{i(d\varphi\theta+\varphi d\theta)}%
=\int(1+id\varphi\theta)\omega e^{i\varphi d\theta}=\int(i\alpha
d\varphi\theta+\beta d\varphi)e^{i\varphi d\theta}=
\]%
\[
=\int(i\alpha\theta+\beta)e^{i\varphi d\theta}=\sum_{n=-\infty}^{\infty}%
\int_{2\pi n}^{2\pi(n+1)}(i\alpha\theta+\beta)e^{i\varphi d\theta}%
=\sum_{n=-\infty}^{\infty}\frac{e^{2\pi(n+1)d\theta}-e^{2\pi(n)d\theta}%
}{id\theta}(i\alpha\theta+\beta)=
\]%
\[
=(i\alpha\theta+\beta)\frac{e^{i2\pi d\theta}-1}{id\theta}\sum_{n=-\infty
}^{\infty}e^{2\pi nd\theta}=(i\alpha\theta+\beta)\frac{e^{i2\pi d\theta}%
-1}{id\theta}\sum_{n=-\infty}^{\infty}\delta(d\theta-n)
\]
Where formal notations like $\frac{f(d\theta)}{d\theta}$ must be interpreted
in the contest of formal power series in $d\theta.$

To check the closure of the class $\tilde{\omega}=i\theta\frac{e^{i2\pi
d\theta}-1}{id\theta}\sum_{n=-\infty}^{\infty}\delta(d\theta-n)$ (for the
other differential form, the closure is trivial) we observe that:
\begin{equation}
d\tilde{\omega}=id\theta\frac{e^{i2\pi d\theta}-1}{id\theta}\sum_{n=-\infty
}^{\infty}\delta(d\theta-n)=\left(  e^{i2\pi d\theta}-1\right)  \sum
_{n=-\infty}^{\infty}\delta(d\theta-n)=0\,. \label{angB}%
\end{equation}

If we take into account the radius $R$ of the circle:
\begin{equation}
\tilde{\omega}=iR\theta\frac{e^{i2\pi Rd\theta}-1}{iRd\theta}\sum_{n=-\infty
}^{\infty}\delta(Rd\theta-n)=
\end{equation}%
\[
=i\theta\frac{1+i2\pi Rd\theta-(2\pi Rd\theta)^{2}+O(d\theta^{2})-1}%
{iRd\theta}\sum_{n=-\infty}^{\infty}\delta(d\theta-n/R)=
\]
\[
=i\theta(2\pi
+O(d\theta))\sum_{n=-\infty}^{\infty}\delta(d\theta-n/R)
\]
In the limit $R\rightarrow\infty$ (flat limit) the series $\sum_{n=-\infty
}^{\infty}\delta(d\theta-n/R)$ gives $\delta(d\theta)$ and therefore the limit
$R\rightarrow\infty$ leads to
\begin{equation}
\lim_{R\rightarrow\infty}\tilde{\omega}=2\pi i\theta\delta(d\theta)\,.
\label{angC}%
\end{equation}
which is the correct Fourier transform of the cohomological class of the flat limit.

\section{d and k differentials}

We now study the image under Fourier transform of the de Rham differential $d$
acting on the complex of differential forms.

If we consider the following diagram:%
\[%
\begin{array}
[c]{ccc}%
\bigwedge^{p}(\mathbb{R}^{n}) & \overset{\mathcal{F}}{\longleftarrow} &
\bigwedge^{n-p}(\mathbb{R}^{n^{\ast}})\\
d\downarrow &  & k\downarrow\\
\bigwedge^{p+1}(\mathbb{R}^{n}) & \overset{\mathcal{F}}{\longrightarrow} &
\bigwedge^{n-p-1}(\mathbb{R}^{n^{\ast}})
\end{array}
\]

the operator $k$ that we want to compute is such that:
\begin{equation}
k=\mathcal{F}\circ d\circ\mathcal{F} \label{koszul}%
\end{equation}
Note that this definition gives $k^{2}=0$, since ${\cal F}^2 = I$ and $d^2 =0$.

We start again with the simple example of $\mathbb{R}^{2}.$ We take $x,y$ as
coordinates in $\mathbb{R}^{2}$ and $u,v$ as dual coordinates in
$\mathbb{R}^{2^{\ast}}.$ We start with the $0-$ forms$.$ In this case
$d\circ \mathcal{F}$ is trivially zero and hence we have that the action of
$k$ on functions is trivial:
\begin{equation}
k\left(  f(u,v)\right)  =0 \label{k0forme}%
\end{equation}

A one form in $\mathbb{R}^{2^{\ast}}$ is $f(u,v)du+g(u,v)dv$ and its
Fourier transform is given by:%
\[
\mathcal{F}\left(  f(u,v)du+g(u,v)dv\right)  =-i\widetilde{f}dy+i\widetilde
{g}dx
\]
The differential is:%
\[
d\left(  -i\widetilde{f}dy+i\widetilde{g}dx\right)  =-i\left(  \frac
{\partial\widetilde{f}}{\partial x}+\frac{\partial\widetilde{g}}{\partial
y}\right)  dxdy=-\left(  \widetilde{uf+vg}\right)  dxdy
\]
Hence we have:%
\begin{equation}
k\left(  f(u,v)du+g(u,v)dv\right)  =\mathcal{F}\left(  -\left(  \widetilde
{uf+vg}\right)  dxdy\right)  =-\left(  uf+vg\right)  \label{k1forme}%
\end{equation}
For the $2-$ forms, written as $f(u,v)dudv,$ we have:
\[
\mathcal{F}\left(  f(u,v)dudv\right)  =\widetilde{f}%
\]
The differential is:%
\[
d\widetilde{f}=\frac{\partial\widetilde{f}}{\partial x}dx+\frac{\partial
\widetilde{f}}{\partial y}dy=-i\widetilde{uf}dx-i\widetilde{vf}dy
\]
Hence:%
\begin{equation}
k\left(  f(u,v)dudv\right)  =\mathcal{F}\left(  -i\widetilde{uf}%
dx-i\widetilde{vf}dy\right)  =-\left(  udv+vdu\right)  f \label{k2forme}%
\end{equation}
The Leibnitz rule is verified:%
\[
k\left(  fdudv\right)  =k\left(  fdu\right)  dv+fdu\, k\left(  dv\right)
\]

The $k$ differential can be computed for generic $n$ and its action on the
functions $f$ and the degree $1$ generators of $\Omega^{\bullet}%
(\mathbb{R}^{n^{\ast}})$ is:%
\begin{align*}
k(f)  &  =0\\
k(fdu^{i})  &  =-u^{i}f
\end{align*}

The differential $k$ was defined here through Fourier transforms but, for
general forms (not only the forms that can be Fourier transformed in some
sense), the (\ref{k0forme}) and (\ref{k1forme}) could be taken as definitions
of the action of a differential operator $k$ on the degree $1$ generators of
$\Omega^{\bullet}(\mathbb{R}^{n^{\ast}})$. The operator is then extended to
$\Omega^{\bullet}(\mathbb{R}^{n^{\ast}})$ using the Leibnitz rule and is a
derivation of degree $-1$. In this broader context the operator just described
is known in mathematics as ``Koszul differential". The formalism of Fourier
transforms can also be used for extending the Koszul differential to the
complexes of super and integral forms.

\section{Picture Changing Operators in QFT}

The Picture Changing Operators (PCO) where introduced in \cite{FMS} in string theory. This is due to the fact that in the quantization of the Ramond-Neveu-Schwarz model
for the fermionic string the sector of superghosts associated to local supersymmetry has an Hilbert space with infinite replicas. Therefore, the vacuum is defined once the
picture is defined and in terms of the vacuum, one can build the vertex operators. However, in amplitude computations one needs
to saturate a certain picture number (depending upon
the moduli of the Riemann surface) and therefore one needs to have vertex operators in different pictures.
The picture  number countd the number
of Dirac delta functions of the superghosts and the PCO can increase or decrease that number at wish.
Notice that the picture number indicates the degree of the form
that can be integrated on a particular Riemann surface.

These operators can also be constructed in our context and they act transversally in the complexes of integral forms. Given a constant commuting vector $v$ we define
the following object
\begin{equation}\label{PCOa}
Y_v = v_\alpha \theta^\alpha \delta(v_\alpha d \theta^\alpha)\,,
\end{equation}
 which has the properties
 \begin{equation}\label{PCOb}
d Y_v = 0\,, Y_v \neq d H\,, ~~~~~  Y_{v + \delta v} = Y_v + d \left(v_\alpha \theta^\alpha \delta_\alpha \theta^\alpha \delta'(v_\alpha d\theta^\alpha) \right)\,,
\end{equation}
where $H$ is an integral form. Notice that $Y_v$ belongs to $\Omega^{(0|1)}$ and by choosing different vectors $v^{(\alpha)}$, we have
\begin{equation}\label{PCOb}
\prod_{\alpha=1}^m Y_{v^{(\alpha)}} = \det( v^{(\alpha)}_\beta ) \theta^{\alpha_1} \dots \theta^{\alpha_m} \delta(d\theta^{\alpha_1}) \dots  \delta(d\theta^{\alpha_m})\,,
\end{equation}
where $v^{(\alpha)}_\beta$ is the $\beta$-component of the $\alpha$-vector. We
can apply the PCO operator on a given integral form by taking the wedge product of the two integral
forms. For example, given $\omega$ in $\Omega^{(p|r)}$ we have
\begin{equation}\label{PCOc}
\omega \longrightarrow \omega\wedge Y_v \in \Omega^{{p|r+1}}\,,
\end{equation}
Notice that if $r =m$, then $ \omega\wedge Y_v =0$; on the other hand, if $v$ does not depend
on the arguments of the delta funtions in $\omega$, then we have a non-vanishing integral form. In addition,
if $d \omega =0$ then $d (\omega \wedge Y_v) =0$ (by applying the Leibniz rule), and if $\omega \neq d \eta$
then it follows that also $\omega \wedge Y_v \neq d U$ where $U$ is an integral form of $\Omega^{(p-1|r+1)}$.
In \cite{Catenacci:2010cs}, it has been proved that $Y_v$ are elements of the de Rham cohomology and
that they are also globally defined. So, given an element of the cohomogy $H_d^{(p|r)}$,
the new integral form $\omega \wedge Y_v$ is an element of $H_d^{(p|r+1)}$.

Let us consider again the example of ${\cal M}^{(2|2)}$ and the 2-form $F = d A \in \Omega^{(2|0)}$
where $A = A_i dx^i + A_\alpha d\theta^\alpha \in \Omega^{(1|0)}$.
Then, we can produce
\begin{equation}\label{PCOd}
F \longrightarrow F \wedge Y_1 \wedge Y_2
\end{equation}
where we have chosen the vector $v^{(1)}$ along the direction
of the first Grassmanian coordinate and $v^{(2)}$ along the other direction. Therefore
we have
\begin{equation}\label{PCOe}
F \wedge Y_1 \wedge Y_2 =
\left(\partial_i A_j dx^i \wedge dx^j + \dots \partial_\alpha A_\beta d\theta^\alpha d\theta^\beta\right) \wedge Y_1 \wedge Y_2
\end{equation}
$$
=( \partial_i A_j \theta^2)  \, dx^i \wedge dx^j  \wedge \delta^2(d\theta) =
( \partial_i A^{(0)}_j \theta^2)  \, dx^i \wedge dx^j  \wedge \delta^2(d\theta)
$$
where $A^{(0)}_j$ is the lowest component of the superfield $A_i$ appearing in the superconnection $A_i$.
The result can be easily integrated in the supermanifold ${\cal M}^{(2|2)}$ yielding the well-known result
\begin{equation}\label{PCOd}
\int_{{\cal M}^{(2|2)}} F \wedge Y_1 \wedge Y_2 = \int \partial_i A^{(0)}_j  dx^i \wedge dx^j
\end{equation}
Since the curvature ${\cal F}^{(2|2)} =F \wedge Y_1 \wedge Y_2$ can be also written as
$d A^{(1|0)} \wedge Y_1 \wedge Y_2$, using that $d Y_i =0$, we have
$$
d \left( A^{(1|0)} \wedge Y_1 \wedge Y_2 \right)  = d {\cal A}^{(1|2)}\,,
$$
where ${\cal A}^{(1|2)}$ is the gauge connection at picture number 2. Notice that performing a gauge transformation on $A$,
we have
$$\delta {\cal A}^{(1|2)} = d \lambda \wedge Y_1 \wedge Y_2 = d \left( \lambda \wedge Y_1 \wedge Y_2 \right)$$
and therefore we can consider $\lambda^{(0|0)} \wedge Y_1 \wedge Y_2$ as the gauge parameter at picture number 2.


\end{document}